\documentclass[10pt, fleqn]{wlscirep}

\usepackage{amsfonts, gensymb, mathtools, microtype}

\usepackage[T1]{fontenc}
\usepackage[utf8]{inputenc}

\usepackage{tikz}
\usetikzlibrary{
    arrows,
    backgrounds,
    calc,
    decorations,
    calligraphy,
    fit,
    positioning,
    shapes,
    spy
}

\usepackage{pgfplots}
\pgfplotsset{compat=newest}
\usepgfplotslibrary{colorbrewer, groupplots}

\usepackage{array, diagbox, makecell, multirow, subfig, xfrac}

\usepackage[export]{adjustbox}

\usepackage{graphicx, graphbox}
\usepackage[figuresright]{rotating}
\graphicspath{{figs/}}

\usepackage{listings}
\usepackage[defaultlines=3, all]{nowidow}

\newcolumntype{P}[1]{>{ \centering  \arraybackslash }p{#1}}
\newcolumntype{Q}[1]{>{ \raggedleft \arraybackslash }p{#1}}
\newcolumntype{M}[1]{>{ \centering  \arraybackslash }m{#1}}
\newcolumntype{N}[1]{>{ \raggedleft \arraybackslash }m{#1}}
\newcolumntype{B}[1]{>{ \centering  \arraybackslash }b{#1}}
\newcolumntype{C}[1]{>{ \raggedleft \arraybackslash }b{#1}}

\usepackage{ccicons}

\usepackage{fontawesome5}

\newcommand{\ORCID}[1]{\textsuperscript{\href{https://orcid.org/#1}{\textcolor[HTML]{A6CE39}{\faOrcid}}}}

\newcommand{\ORCIDSchlosser}{0000-0002-0682-4284} 
\newcommand{\ORCIDBeuth}{0000-0001-5482-9787}     
\newcommand{\ORCIDMeyer}{0000-0002-3372-1619}     
\newcommand{\ORCIDKumar}{0000-0002-2654-7050}     
\newcommand{\ORCIDFurashova}{0000-0002-1273-7005} 
\newcommand{\ORCIDKowerko}{0000-0002-4538-7814}   

\definecolor{custom_blue}{RGB}{0, 114, 189}

\begin{document}

\title{Visual Acuity Prediction on Real-Life Patient Data Using a Machine Learning Based Multistage System}

\author[1,\textasteriskcentered]{Tobias Schlosser\ORCID{\ORCIDSchlosser}}
\author[1]{Frederik Beuth\ORCID{\ORCIDBeuth}}
\author[1]{Trixy Meyer\ORCID{\ORCIDMeyer}}
\author[1]{Arunodhayan Sampath Kumar\ORCID{\ORCIDKumar}}
\author[2]{Gabriel Stolze}
\author[2]{Olga Furashova\ORCID{\ORCIDFurashova}}
\author[2]{Katrin Engelmann}
\author[1,\textasteriskcentered]{Danny Kowerko\ORCID{\ORCIDKowerko}}

\affil[1]{
    Junior Professorship of Media Computing,
    Chemnitz University of Technology,
    09107 Chemnitz, Germany
}
\affil[2]{
    Department of Ophthalmology,
    Klinikum Chemnitz gGmbH,
    09116 Chemnitz, Germany}
\affil[$\text{\textasteriskcentered}$]{
    Correspondence: tobias.schlosser@cs.tu-chemnitz.de, danny.kowerko@cs.tu-chemnitz.de
}

\begin{abstract}
    In ophthalmology, intravitreal operative medication therapy (IVOM) is a widespread treatment for diseases related to the age-related macular degeneration (AMD), the diabetic macular edema (DME), as well as the retinal vein occlusion (RVO). However, in real-world settings, patients often suffer from loss of vision on time scales of years despite therapy, whereas the prediction of the visual acuity (VA) and the earliest possible detection of deterioration under real-life conditions is challenging due to heterogeneous and incomplete data. In this contribution, we present a workflow for the development of a research-compatible data corpus fusing different IT systems of the department of ophthalmology of a German maximum care hospital. The extensive data corpus allows predictive statements of the expected progression of a patient and his or her VA in each of the three diseases. For the disease AMD, we found out a significant deterioration of the visual acuity over time. Within our proposed multistage system, we subsequently classify the VA progression into the three groups of therapy ``winners'', ``stabilizers'', and ``losers'' (WSL classification scheme). Our OCT biomarker classification using an ensemble of deep neural networks results in a classification accuracy (F1-score) of over $98$~\%, enabling us to complete incomplete OCT documentations while allowing us to exploit them for a more precise VA modeling process. Our VA prediction requires at least four VA examinations and optionally OCT biomarkers from the same time period to predict the VA progression within a forecasted time frame, whereas our prediction is currently restricted to IVOM / no therapy. We achieve a final prediction accuracy of $69$~\% in macro average F1-score, while being in the same range as the ophthalmologists with $57.8$ and $50 \pm 10.7$~\% F1-score.
\end{abstract}


\maketitle

\section{Introduction and motivation}
\label{section:introduction_and_motivation}

High-resolution imaging of the central retina utilizing optical coherence tomography (OCT) plays a key role in the diagnosis and monitoring of the most common macular diseases such as age-related macular degeneration (AMD), diabetic macular edema (DME), and retinal vein occlusion (RVO) \cite{Nimse2016, Seeboeck2019}. OCT biomarkers are specific properties of measurements that are extracted from the OCT images to provide information about the condition of the tissues and tissue layers within the human eye. Furthermore, detailed analysis of different biomarkers on OCT scans is now the basis for treatment decisions as several biological markers provide not only information on diagnosis of these particular eye conditions, but also play an important role in predicting the treatment response. With the increasing amount of available data for therapeutic strategies, the identification of biomarkers with predictive values, as well as different medication options such as intravitreal operative medication therapies (IVOM) for macular edema or diabetic retinopathy patients, it is challenging for ophthalmologists to individualize the therapy for each patient. Artificial intelligence (AI) based algorithms should, in the future, help to find optimal individual therapeutic strategies for each patient. Real-world studies have already demonstrated how important the early treatment begin and the compliance with the therapy are \cite{Finger2013, Hykin2016, Wintergerst2019}. Yet, real-world studies need more attention, where we can contribute with a settings of real-world study data.
%
The study group of \textit{Gerding et al.} was one of the first ones to classify patients with AMD into three treatment success groups based on visual acuity (VA) and central retinal thickness: therapy ``winners'', ``stabilizers'', and ``losers'' (WSL classification scheme) \cite{Gerding2011}. This interdisciplinary cooperation between IT specialists and ophthalmologists aims at analyzing the patients' data according to the WSL classification scheme while identifying predictive values for several OCT biomarkers. The results should help ophthalmologists to better define their therapy strategy for each patient in everyday practice. Moreover, \textit{Schmidt-Erfurth et al.} recently reported the potential of AI-based approaches for targeted optimization of diagnosis and therapy for eye diseases \cite{Gerendas2018}. In their contributions, they furthermore describe the impact of deep learning (DL) for the prediction of patient progressions in the earlier stages of AMD utilizing OCT biomarkers \cite{Schmidt-Erfurth2020, Waldstein2020}. Whereas current state-of-the-art research also explores the explainability as well as the related nomenclature when reporting AMD-related diseases \cite{Spaide2020, Matsui2022}, AMD such as exudative AMD as well as DME and RVO can be seen as the three most prevalent investigated eye diseases within the context of AI \cite{Chen2021, Phan2021, Romond2021, Matsui2022}.

\begin{figure}[tb]
    \centering

    \includegraphics[width=0.99\textwidth, trim={0 12cm 0 0}, clip, frame]{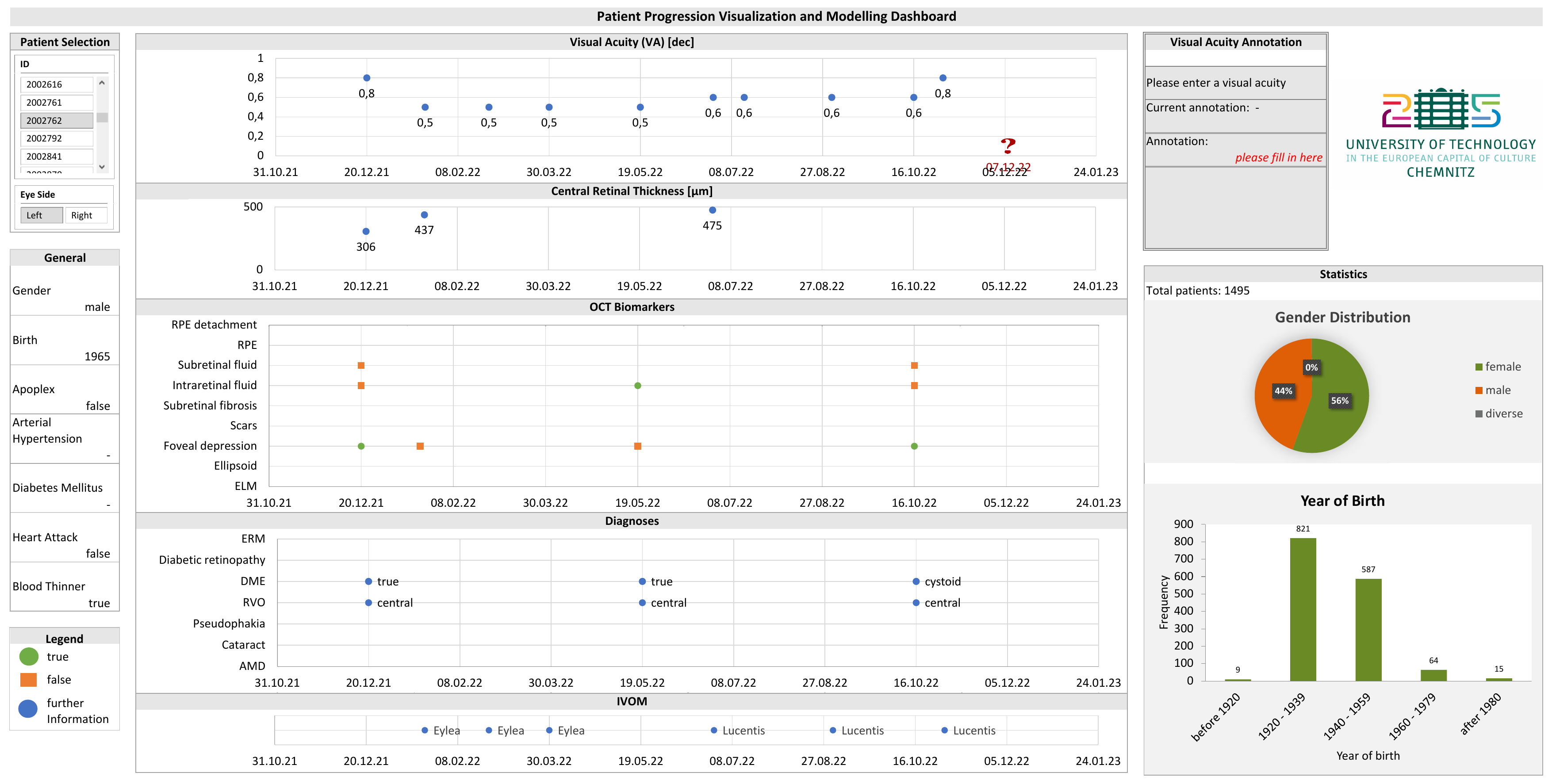}

    \caption{Patient progression visualization and modeling dashboard developed with medical doctors for medical doctors as well as researchers. Visualized are general patient information, visual acuity, OCT biomarkers, diagnoses, and medications. It is possible to annotate the expected course of the VA on site at the position of the red question mark. The shown data set is inspired by real patients and is synthesized to avoid re-identification. The distance between two adjacent vertical guide lines is 50 days.}
    \label{figure:dashboard}
\end{figure}

\subsection{Related work}
\label{section:related_work}

In the following sections, we discuss ophthalmic research on data mining from clinical IT systems (section~\ref{section:availability_of_ophthalmic_data_in_research}), text and OCT image processing (section~\ref{section:ophthalmic_data_processing}), and the use of the processed data for patient progression modeling (section~\ref{section:patient_progressing_modeling}).

\subsubsection{Availability of ophthalmic data in research}
\label{section:availability_of_ophthalmic_data_in_research}

In recent years, more and more ophthalmic data sets are being released \cite{Porwal2018, Pachade2021}. A recent review article identified 94 open access data sets containing 507\,724 images and 125 videos from 122\,364 patients with diabetic retinopathy, glaucoma, and AMD being disproportionately over-represented in comparison to other eye diseases. However, the documentation of demographic characteristics such as age, sex, and ethnicity was reported to be of poor data quality even at the aggregate level \cite{Khanani2020}.
%
In 2017, we proposed a prototypical workflow to aggregate ophthalmic text and image data of all patients from the Department of Ophthalmology of the maximum care hospital Klinikum Chemnitz gGmbH in Chemnitz, Germany. We combined data mining and basic natural language (NLP) processing utilizing the interface of the clinic's practice management software Turbomed (CompuGroup Medical) and extracted a set of preliminary diagnostic patient data in order to determine the ratio of patients with VA improvement, stabilization, and deterioration \cite{Roessner2017}.

\subsubsection{Ophthalmic text and OCT image processing}
\label{section:ophthalmic_data_processing}

While widely being used for general text processing, NLP systems have recently been demonstrated to robustly extract medication information from clinical notes to study VA, intraocular pressure, and medication outcomes of cataract and glaucoma surgeries \cite{Xu2010, Wang2020} to develop predictive models for low-vision prognosis \cite{Wang2021} as well as to predict glaucoma progressions \cite{Hu2022}.
%
Following \textit{De Fauw et al.}, especially machine learning (ML) and deep learning based approaches enable a more precise progression modeling as recent advances prove their applicability and capabilities within the domain \cite{De_Fauw2018}. Moreover, recognition of OCT biomarkers \cite{Kurmann2019} allows further VA and treatment based medical forecasts, including ensemble-based solutions to OCT segmentation \cite{Gorgi2017} that enable the completion of incomplete OCT documentations. However, automated thresholding algorithms can yield an improved reproducibility of OCT parameters while allowing a more sensitive diagnosis of pathologies when, e.g., discriminating between healthy and impaired macular perfusion \cite{Terheyden2020}.

\begin{figure}[tb]
    \centering

    \begin{tikzpicture}[
     d/.style={draw, diamond},
     e/.style={draw, ellipse},
     p/.style={draw, trapezium, trapezium left angle=75, trapezium right angle=105},
     r/.style={draw, rectangle, minimum height=0.55cm}]
        \node[r, minimum width=3cm] (l1n1) at (-4, 0) {EHR system};
        \node[r, minimum width=3cm] (l1n2) at ( 0, 0) {CIS system};
        \node[r, minimum width=3cm] (l1n3) at ( 4, 0) {OCT system};

        \node[r, minimum width=3cm, align=center] (l2n1) at (-4, -1.5) {BDT-based data  \\ extraction \\ (*.BDT interface)};
        \node[r, minimum width=3cm, align=center] (l2n2) at ( 0, -1.5) {IVOM-based data \\ extraction \\ (*.CSV interface)};
        \node[r, minimum width=3cm, align=center] (l2n3) at ( 4, -1.5) {E2E-based data  \\ extraction \\ (*.E2E interface)};

        \node[r, minimum width=4cm] (l3n1) at (0, -3) {Patient-based data merging};
        \node[r, minimum width=4cm] (l3n2) at (6, -3) {Dashboard visualization};
        \node[r, minimum width=4cm, draw=none] (l3n0) at (-6, -3) {};

        \node[r, minimum width=3cm] (l4n1) at (-4, -4.5) {Anamnesis};
        \node[r, minimum width=3cm] (l5n1) at (-4, -5.5) {Disease};
        \node[r, minimum width=3cm] (l6n1) at (-4, -6.5) {IVOM};
        \node[r, minimum width=3cm] (l7n1) at (-4, -7.5) {Visual acuity};
        \node[draw, inner sep=0.2cm, fit=(l4n1) (l5n1) (l6n1) (l7n1)] (f1) {};
        \node[above] (f1l) at (f1.north) {Category-centered tables};

        \node[r, minimum width=3cm] (l4n2) at (0, -4.5) {Global statistics};
        \node[r, minimum width=3cm] (l5n2) at (0, -5.5) {Anamnesis statistics};
        \node[r, minimum width=3cm] (l6n2) at (0, -6.5) {Disease statistics};
        \node[r, minimum width=3cm] (l7n2) at (0, -7.5) {IVOM statistics};
        \node[draw, inner sep=0.2cm, fit=(l4n2) (l5n2) (l6n2) (l7n2)] (f2) {};
        \node[above] (f2l) at (f2.north) {Predictive statistics};

        \draw[->] (l4n1.east) -- (l5n2.west);
        \draw[->] (l5n1.east) -- (l6n2.west);
        \draw[->] (l6n1.east) -- (l7n2.west);

        \node[r, minimum width=3cm] (l4n3) at (4, -4.5) {Local OCT class.};
        \node[r, minimum width=3cm] (l5n3) at (4, -5.5) {Global OCT class.};
        \node[r, minimum width=3cm] (l6n3) at (4, -6.5) {VA prediction};
        \node[r, minimum width=3cm] (l7n3) at (4, -7.5) {Treatment adjustment};
        \node[draw, inner sep=0.2cm, fit=(l4n3) (l5n3) (l6n3) (l7n3)] (f3) {};
        \node[above] (f3l) at (f3.north) {Classification and prediction};

        \draw[->] (l4n3) -- (l5n3);
        \draw[->] (l5n3) -- (l6n3);

        \node[r, minimum width=4cm] (l8n1) at (0, -9) {Treatment recommendation};

        \draw[->] (l1n1) -- (l2n1);
        \draw[->] (l1n2) -- (l2n2);
        \draw[->] (l1n3) -- (l2n3);

        \draw[->] (l2n1) -- (l3n1);
        \draw[->] (l2n2) -- (l3n1);
        \draw[->] (l2n3) -- (l3n1);

        \draw[->] (l3n1) -- (l3n2);
        \draw[->] (l3n1) -- (f1l.north);
        \draw[->] (l3n1) -- (f2l.north);
        \draw[->] (l3n1) -- (f3l.north);

        \draw[->] (f1.south) -- (l8n1);
        \draw[->] (f3.south) -- (l8n1);
    \end{tikzpicture}

    \caption{Proposed medical text and image data mining workflow for the following descriptive statistics and VA modeling based on our three base systems of the electronic health record system (EHR or Turbomed system), the clinical information system (CIS system), and the OCT system. Our patient progression visualization and modeling dashboard is shown in Fig.~\ref{figure:dashboard}. Currently, the classification and prediction step of treatment adjustment is still in development. However, given all data, a treatment recommendation can be realized.}
    \label{figure:workflow}
\end{figure}
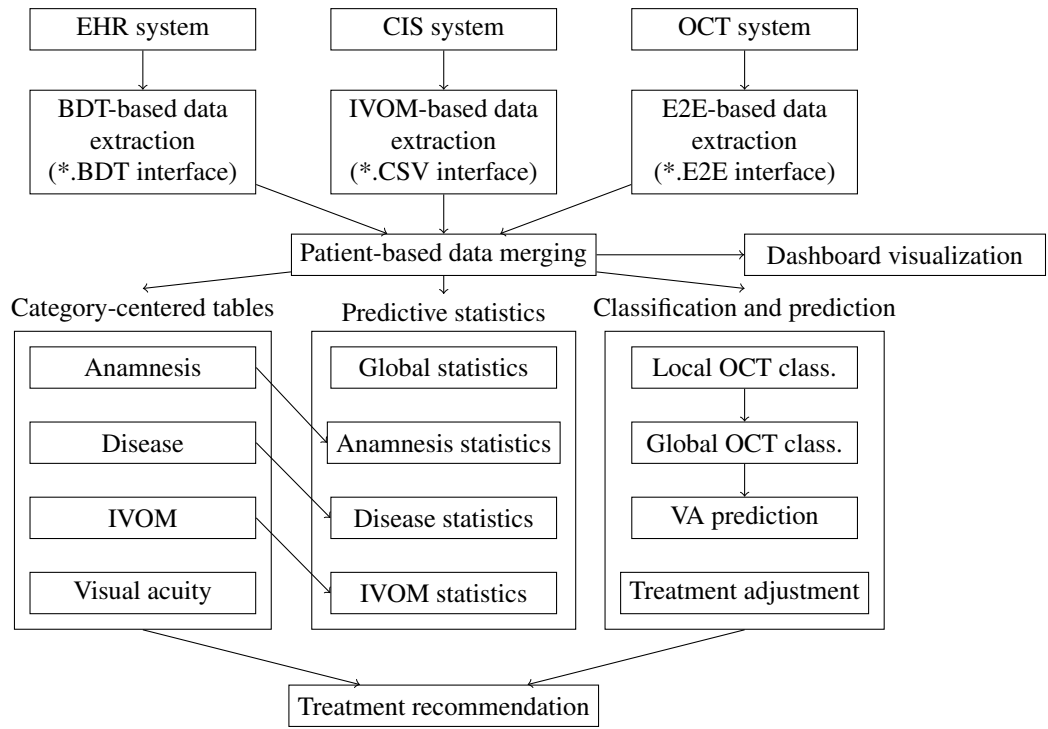

\subsubsection{Patient progressing modeling}
\label{section:patient_progressing_modeling}

\textit{Schmidt-Erfurth et al.} investigate the influence of hyperreflective foci as OCT biomarkers during the progression of geographic atrophy within the context of AMD \cite{Schmidt-Erfurth2020}. While utilizing deep neural networks (DNN) for OCT segmentation \cite{Schlegl2018}, they identify and localize occurrences given a data set of 87 eyes from 54 different patients. Following \textit{Schmidt-Erfurth et al.'s} contribution, \textit{Waldstein et al.} propose a further developed system while evaluating it with 8\,529 OCT volumes of 512 different patients \cite{Waldstein2020}. Moreover, time-dependent sequence modeling using recurrent neural networks (RNN) \cite{Rumelhart1986} constitutes a promising approach to treatment prediction \cite{Wang2018}. At this point it is also noted that approaches to patient progression modeling exist that explore conventional models. These include mathematical models, e.g., to determine the effect of the anti-vascular endothelial growth factor on VA via medication concentration and tolerance \cite{Edwards2020}, as well as regression-based approaches, e.g., to predict VA in diabetic retinopathy via the foveal avascular zone area \cite{DaCosta2020}.

\subsection{Contribution of this work}
\label{section:contribution_of_this_work}

In this contribution, we present an IT system architecture that aggregates patient information for more than 49\,000 patients from different categories of various multimedia data in the form of text and images within multiple heterogeneous ophthalmic data resources.
%
The resulting data corpus enables predictive statements to be made about the expected progression of a patient's visual acuity after at least four VA examinations in each of the three diseases~-- AMD, DME, and RVO. A more fine-grained analysis is conducted to reveal the influence of medical co-existing factors such as other diseases in this real-world setting.
%
Within our proposed multistage system, an ensemble of deep neural networks allows the completion of incomplete or missing OCT documentations. In order to conduct a patient progression modeling, we define a fundamental use case for the prediction of the VA by aggregating different patient information as input of the subsequent VA forecast after a given time period. We present an evaluation formalism and discuss the resulting predictions in comparison to those of a human annotator such as experienced ophthalmologists.
%
In order to enable ophthalmic doctors to annotate their predictions regarding the patient-wise expected VA progression, our proposed patient progression visualization and modeling dashboard gives an overview of our aggregated data with patient-wise information of, i.a., general patient information and VA, OCT biomarkers, diagnosis, and medication information (Fig.~\ref{figure:dashboard}).

\begin{table}[tb]
    \centering

    \begin{tabular}{|c|c|}
        \hline
        Disease & Rules \\
        \hline
        \noalign{\vskip 2pt}

        \hline
        AMD &
        \begin{tabular}{ccc}
            \textit{`feucht', `feuchte', `exsudativ', `exsudative', *} & $\rightarrow$ & \textit{True} \\
            \textit{`trocken', `trockene'}                             & $\rightarrow$ & \textit{False}
        \end{tabular} \\
        \hline

        DME &
        \begin{tabular}{ccc}
            \textit{`diabetisch', `diabetisches'} & $\rightarrow$ & \textit{True} \\
            \textit{*}                            & $\rightarrow$ & \textit{False}
        \end{tabular} \\
        \hline

        RVO &
        \begin{tabular}{ccc}
            \textit{`ast', `retinal', `zentral'} & $\rightarrow$ & \textit{True} \\
            \textit{*}                           & $\rightarrow$ & \textit{False}
        \end{tabular} \\
    \hline
    \end{tabular}

    \caption{Exemplary (German) ophthalmic text processing rules for exudative AMD, DME, and RVO.}
    \label{table:text_processing_rules}
\end{table}

\section{Methods and implementation}
\label{section:methods_and_implementation}

This section provides the methods and their implementation related to the patient progression modeling. Firstly, the proposed system's architecture for data acquisition, preprocessing, analysis, and prediction is introduced (section~\ref{section:system_architecture}). To allow a unified data processing, the role of the related terminologies for medical application (ophthalmic ontology) is addressed (section~\ref{section:text_processing_and_ontology}), whereas the data fusion and cleaning is introduced (section~\ref{section:data_fusion_and_cleaning}). Subsequently, descriptive statistics are derived (section~\ref{section:descriptive_statistics}). Finally, the principles of the patient progression modeling within the context of our ML- and DL-based approaches are introduced (section~\ref{section:patient_progression_modeling}).

The related implemented models and approaches to patient progression modeling as well as OCT biomarker classification are provided via open science with the machine learning framework Hexnet \cite{Schlosser2019_ICMLA} and can be found on its project page and repository under \url{https://github.com/TSchlosser13/Hexnet} (see also \texttt{\_ML/models/contrib}).
%
The framework Hexnet provides the functionality that mainly allows for the utilization of out-of-the-box ML- and DL-based methods and models, including common routines for data storage management, preprocessing, model training and testing, as well as evaluation. It was originally developed for hexagonal image processing and deep learning, while it has been recently further developed for classical machine learning. Its machine learning module (directory \texttt{\_ML/}) is based on the machine learning library TensorFlow with its front end Keras, whereas scikit-learn is deployed for all machine learning based models and evaluation procedures. Within the current research work, Hexnet's machine learning module has been extended to enable the processing of ophthalmic imagery, the data handling of our data vectors, as well as our ophthalmic evaluation through models such as, e.g., statistical, moving average (MA), and weighted MA estimators as well as recurrent neural networks (RNN). These new contributions can be found under \texttt{models/contrib/} via \texttt{ophthalmology\_evaluation.py} and \texttt{RNNs.py}). In comparison, already existing ML and DL models are also present within \texttt{models/} and \texttt{models/contrib/}, covering regressors and our multilayer perceptron approaches as well as DenseNet201 and ResNet152V2 that are based on Keras or scikit-learn.

The present study was approved by the Institutional Review Board of Saxony (Dresden, Germany) under the number EK-BR-102/20-2. We confirm that all research was performed in accordance with relevant guidelines and regulations. Informed patients' consent was waived because of the retrospective anonymous design and because no study-related investigations were necessary.

\subsection{System architecture}
\label{section:system_architecture}

Our ophthalmic core data set obtained from data as well as text mining is based on the categories of general patient information (\textit{G}), VA-based patient information (\textit{V}), OCT scans and biomarkers (\textit{O}), diseases (\textit{D}), as well as treatments and medications (\textit{T}). These categories are obtained from different base systems, the electronic health record system (EHR system, Turbomed), the clinical information system (CIS system, SAP), and an OCT system (Heidelberg Eye Explorer) (see also Fig.~\ref{figure:workflow}). While all three systems contain basic patient information, the OCT system provides first and foremost OCT scans (categories \textit{G} and \textit{O}). The EHR system consists of one large database which contains all relevant patient information within the categories of \textit{G}, \textit{V}, \textit{D}, and \textit{T}. As the treatment and medication information within the EHR system is not always complete, the clinic's CIS system is additionally utilized in order to retrieve missing medication and therapy information (category \textit{CIS}). Following the retrieved data of all three base systems in the form of BDT (EHR system), CSV (CIS system), and E2E files (OCT system), all relevant information are merged in a patient-centered way. This includes a chronological synchronization of all patient information, whereas sensitive patient information has to be furthermore pseudo-anonymized due to patient data privacy and protection laws. These results are then presented via our patient progression visualization and modeling dashboard (Fig.~\ref{figure:dashboard}).

\begin{figure}[tbp]
    \definecolor{LimeGreen}{HTML}{8DC73E}
    \definecolor{Red}{HTML}{ED135A}

    \centering

    \begin{tikzpicture}
        \begin{groupplot}[
         group style={
             group size=1 by 3,
             vertical sep=2pt,
             xlabels at=edge bottom,
             xticklabels at=edge bottom
         },
         width=0.7\textwidth, height=0.33\textwidth,
         xlabel={VA measurement number $i$},
         y label style={at={(axis description cs:-0.1,0.5)}},
         yticklabel style={/pgf/number format/.cd, precision=1, fixed, fixed zerofill},
         legend pos=outer north east, legend cell align=left,
         xmin=-1, xmax=19]
            \nextgroupplot[ylabel={VA [logMAR]}, ymin=0.45, ymax=1.05, y dir=reverse, ytick={0.5, 0.6, 0.7, 0.8, 0.9, 1.0}]

            \addplot[mark=square*, mark options={draw=black!80, fill=custom_blue}] coordinates {
                (0, 0.6) (1, 0.6) (2, 0.8) (3, 0.5) (4, 0.8) (5, 0.5) (6, 0.5) (7, 0.5) (8, 0.5) (9, 0.6) (10, 0.6) (11, 0.6) (12, 0.6) (13, 0.8) (14, 0.8) (15, 0.8) (16, 0.8) (17, 0.8) (18, 1.0)
            };
            \addlegendentry{Ground truth}

            \addplot[mark=*, mark options={scale=1.75, draw=black!80, fill=custom_blue}] coordinates {
                (4, 0.6) (5, 0.6) (6, 0.6) (7, 0.55) (8, 0.55) (9, 0.55) (10, 0.6) (11, 0.6) (12, 0.6) (13, 0.75) (14, 0.75) (15, 0.75) (16, 0.75) (17, 0.75) (18, 0.75)
            };
            \addlegendentry{Predictions model}

            \addplot[mark=o, mark options={scale=1.75, draw=black!80}] coordinates {
                (4, 0.5) (5, 0.5) (6, 0.5) (7, 0.5) (8, 0.5) (9, 0.6) (10, 0.6) (11, 0.6) (12, 0.6) (13, 0.8) (14, 0.8) (15, 0.8) (16, 0.8) (17, 0.8) (18, 0.8)
            };
            \addlegendentry{Predictions ophthalmologist}

            \nextgroupplot[ylabel={$\Delta\text{VA [logMAR]}$}, ymin=-0.35, ymax=0.35, y dir=reverse, ytick={-0.3, -0.2, -0.1, 0.0, 0.1, 0.2, 0.3}, extra y ticks={-0.1, 0.1}, extra y tick labels={}, extra y tick style={grid=major}]

            \node at (axis cs:1,-0.2) {Winners};
            \node at (axis cs:1,-0.0) {Stabilizers};
            \node at (axis cs:1, 0.2) {Losers};

            \addplot[mark=square*, mark options={draw=black!80, fill=custom_blue}] coordinates {
                (4, 0.3) (5, -0.3) (6, 0.0) (7, 0.0) (8, 0.0) (9, 0.1) (10, 0.0) (11, 0.0) (12, 0.0) (13, 0.2) (14, 0.0) (15, 0.0) (16, 0.0) (17, 0.0) (18, 0.2)
            };
            \addlegendentry{Ground truth}

            \addplot[mark=*, mark options={scale=1.75, draw=black!80, fill=custom_blue}] coordinates {
                (4, 0.0) (5, 0.0) (6, 0.0) (7, -0.05) (8, 0.0) (9, 0.0) (10, 0.05) (11, 0.0) (12, 0.0) (13, 0.15) (14, 0.0) (15, 0.0) (16, 0.0) (17, 0.0) (18, 0.0)
            };
            \addlegendentry{Predictions model}

            \addplot[mark=o, mark options={scale=1.75, draw=black!80}] coordinates {
                (4, 0.0) (5, 0.0) (6, 0.0) (7, 0.0) (8, 0.0) (9, 0.1) (10, 0.0) (11, 0.0) (12, 0.0) (13, 0.2) (14, 0.0) (15, 0.0) (16, 0.0) (17, 0.0) (18, 0.0)
            };
            \addlegendentry{Predictions ophthalmologist}

            \nextgroupplot[ylabel={WSL prediction}, ymin=-0.5, ymax=2.5, ytick={0, 1, 2}, yticklabels={L, S, W}]

            \addplot[only marks, mark=*, mark options={scale=1.75, draw=black!80, fill=LimeGreen}, y filter/.expression=y-0.125] coordinates {
                (6, 1) (7, 1) (8, 1) (9, 1) (10, 1) (11, 1) (12, 1) (13, 0) (14, 1) (15, 1) (16, 1) (17, 1)
            };
            \addlegendentry{Correct prediction: model}

            \addplot[only marks, mark=o, mark options={scale=1.75, draw=LimeGreen}, y filter/.expression=y+0.125] coordinates {
                (6, 1) (7, 1) (8, 1) (9, 1) (10, 1) (11, 1) (12, 1) (13, 0) (14, 1) (15, 1) (16, 1) (17, 1)
            };
            \addlegendentry{Correct prediction: ophthalm.}

            \addplot[only marks, mark=*, mark options={scale=1.25, draw=black!80, fill=Red}, y filter/.expression=y-0.125] coordinates {
                (4, 1) (5, 1) (18, 1)
            };
            \addlegendentry{Incorrect prediction: model}

            \addplot[only marks, mark=o, mark options={scale=1.25, draw=Red}, y filter/.expression=y+0.125] coordinates {
                (4, 1) (5, 1) (18, 1)
            };
            \addlegendentry{Incorrect prediction: ophthalm.}
        \end{groupplot}
    \end{tikzpicture}

    \caption{Visual acuity (VA) modeling and evaluation principles. Given a set of past visual acuity examinations, we predict the visual acuity value at the next time step. \textbf{Top:} Visual acuity progression of our exemplary patient from Fig.~\ref{figure:dashboard}. Predictions of the visual acuity are conducted by our MLP-LDA model starting with the 5th VA value. In addition, we asked an ophthalmologist to predict the VA. As shown in our visualization, both the predictions the ophthalmologist as well as our proposed model lie close to each other, which in turn emphasizes their prediction capabilities within the context of our ground truth data. \textbf{Middle:} Difference of the visual acuity from the previous to the current data point, showing the improvement, stabilization, or aggravation of the VA from the last examination. A threshold of $0.1$ logMAR units divides the patient into three groups: $\Delta\textit{VA}_i < -0.1$ into winners (W), $-0.1 \leq \Delta\textit{VA}_i \leq 0.1$ into stabilizers (S), and $\Delta\textit{VA}_i > 0.1$ into losers (L), whereby $i$ denotes the data point with $\Delta\textit{VA}_i = \textit{VA}_i - \textit{VA}_{i - 1}$. The first four data points of the ground truth are omitted within the visualization. \textbf{Bottom:} Prediction quality of our model and the ophthalmologist. Shown are the predictions of the patient for our approach to VA prediction and the ophthalmologist, whereas correct (\textcolor{green}{\textbullet}) and incorrect (\textcolor{red}{\textbullet}) WSL predictions are highlighted. The underlying ophthalmic feature vector data organization for VA prediction is depicted in Table~\ref{table:VA_modeling_dataset_overview}.}
    \label{figure:VA_modeling}
\end{figure}
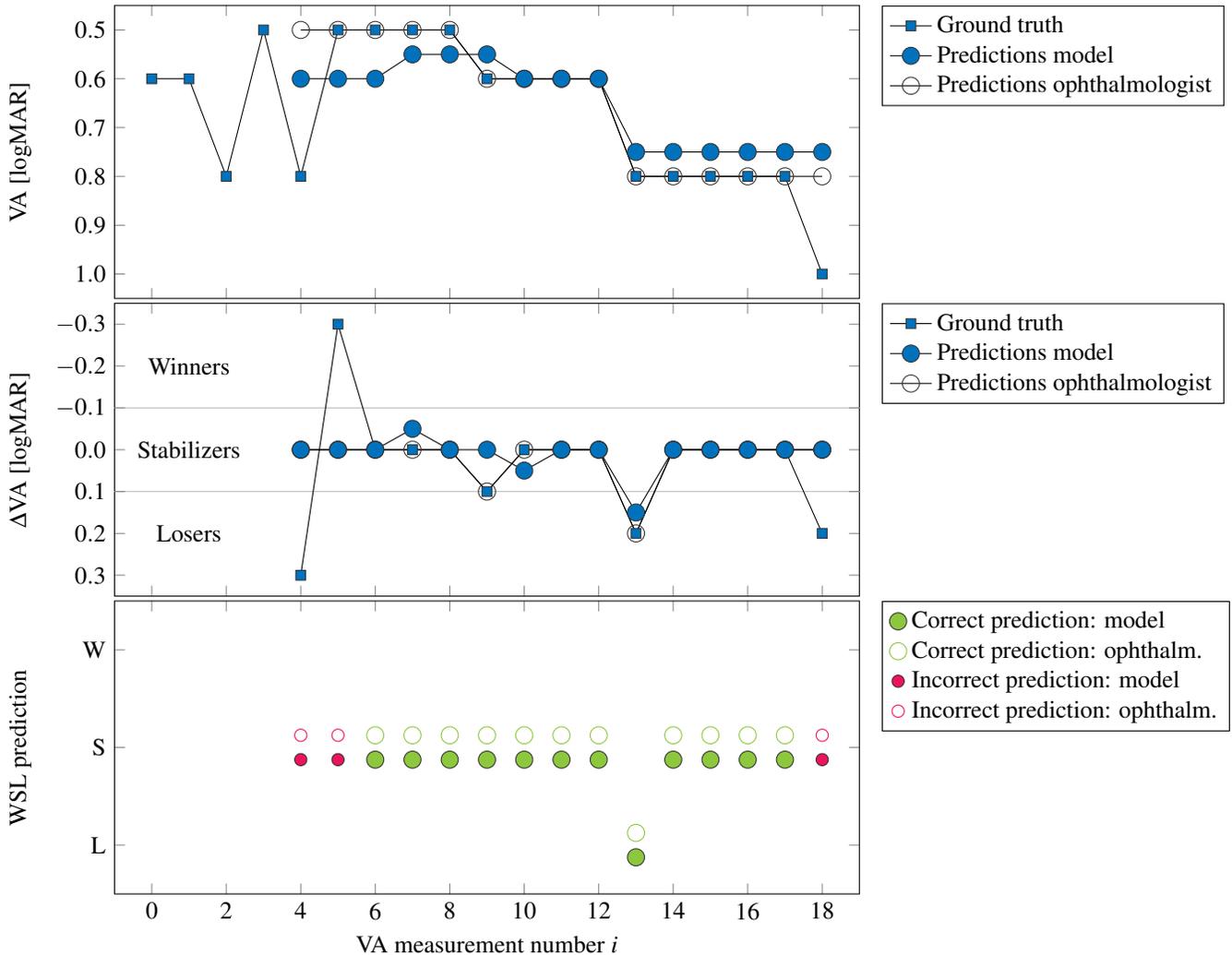

\subsection{Text processing and ontology}
\label{section:text_processing_and_ontology}

The challenge of the ophthalmic text processing is given by the heterogeneity and incompleteness of the underlying data itself, which are in turn documented by many different medical doctors. For text processing and ontology creation, the programming language Python with the Natural Language Toolkit (NLTK) \cite{Bird2009} was utilized.

In order to harmonize our weakly-structured medical texts present within our different databases extracted from electronic health records from over 10 years ranging from 2010 to 2020, we applied a general stemming approach using a snowball stemmer for the German language, also known as the Porter stemming algorithm \cite{porter1980algorithm}. Within computer linguistics, this approach enables an automatic tracing and finally a reduction of words to their stem or root word to allow for a unified text representation. To this end, a set of rules is applied until the current word's stem is extracted \cite{porter1980algorithm}, for which NLTK provides our language-specific rules. When processing, the stemmer is applied iteratively to every word given within the current medical text. Following the obtained stems, specific ophthalmic and medical rules are applied for further processing.

These rules encompass a set of category-specific rules handling abbreviations, negations, and synonyms as well as orthography and grammar variants and mistakes. For this purpose, recognized medical and ophthalmic text strings are mapped into a unique ontology. In the following, we demonstrate some arbitrary strings from the EHR system's diagnosis with customized (German) abbreviations and how they are mapped to diseases, whereby ``$*$'' denotes a placeholder for any string that is not handled otherwise. The mapping is case-insensitive and was specifically adjusted for German doctors (Table~\ref{table:text_processing_rules}). The results have been qualitatively inspected for reliability and plausibility. A thorough quantitative evaluation of our text processing algorithms is still subject to further investigations and will be part of our future research. Therefore, it is beyond the scope of this contribution to further explain the underlying (German) text processing in detail.

\begin{table}[tb]
    \centering

    \begin{tabular}{|M{3cm}|M{2cm}|M{2cm}|M{2cm}|}
        \cline{2-3}
        \multicolumn{1}{c}{} & \multicolumn{2}{|c|}{Class} & \multicolumn{1}{c}{} \\
        \cline{2-3}
        \noalign{\vskip 2pt}

        \hline
        OCT biomarker & Physiological & Pathological & Class ratio [$\sfrac{\text{phys.}}{\text{path.}}$] \\
        \hline
        \noalign{\vskip 2pt}

        \hline
        ELM                 & $\hphantom{0}7\,660$ & $\hphantom{00\,}407$ & $18.8$ \\
        \hline
        Ellipsoid           & $\hphantom{0}7\,377$ & $\hphantom{00\,}308$ & $24.0$ \\
        \hline
        Foveal depression   & $40\,087$ & $10\,923$ & $\hphantom{0}3.7$ \\
        \hline
        RPE                 & $\hphantom{0}7\,866$ & $10\,635$ & $\hphantom{0}0.7$ \\
        \hline
        Scars               & $\hphantom{0}9\,088$ & $\hphantom{0}4\,687$ & $\hphantom{0}1.9$ \\
        \hline
        Subretinal fibrosis & $\hphantom{0}8\,481$ & $\hphantom{0}5\,261$ & $\hphantom{0}1.6$ \\
        \hline
    \end{tabular}

    \caption{Ophthalmic data set overview of our 6 different OCT biomarkers. Listed are the available OCT slices per OCT biomarker with their related classes of physiological as well as pathological image samples.}
    \label{table:OCT_biomarker_dataset_overview}
\end{table}

\subsection{Data fusion and cleaning}
\label{section:data_fusion_and_cleaning}

Our practice management software supports database exports via the BDT file format. Here, we extracted electronic health records from over 10 years ranging from 2010 to 2020, including over 49\,000 patients and more than 130\,000 examinations. Our six main categories (G, V, O, D, T, and OP) currently span a total of more than 30 subcategories. These include, among others, medication-related information such as information on apoplexy and blood thinning as well as biomarker-related information such as information on central retinal thickness and intraretinal fluid. More than 18\,000 IVOMs are available from 2013 to 2020 after a fusion of the EHR and CIS systems' data. All data were linked to the over 12\,000 OCT volumes exported from the OCT system via the patient ID (see also Fig.~\ref{figure:workflow}, patient-based data merging).

Subsequently, the obtained merged data table is further processed and cleaned. Firstly, it is cleaned by mapping the arbitrary text strings present in the medical letters to diseases. The medical letters contain several terms and abbreviations for specific diseases as different letters stem from several different doctors (see previous section, Table~\ref{table:text_processing_rules}). Secondly, unspecific and invalid terms are revised. For example, the entry ``eye side'' of treatments is only allowed to have the entries ``left'' or ``right'', while invalid entries such as ``-'' are removed.

\subsection{Category-centered data organization and descriptive statistics}
\label{section:descriptive_statistics}

After the data fusion and cleaning, the data are focused towards the description of (i) anamnesis, (ii) intravitreal operative medications (IVOMs), (iii) diseases, as well as (iv) visual acuity (Fig.~\ref{figure:workflow}). The merged main table is changed towards a patient-centered description. The aim is to record the start and the end dates of anamnesis entries (if available), the diseases, and the IVOM therapy cycles for each patient. We designed our tables via two lines per patient since the diseases are to a large degree eye independent, for which we include both eyes as separate lines in our tables. For the medical doctors, the category-centered tables allow an easier filtering, an easier sorting, and a more compact view of the data for a particular medical information.

Finally, all tables are visualized in the statistical-description module that illustrates for example a single data set. In addition, the tables can also be combined to show cross-table-referenced data correlations. From all available visualizations, including up to 30 combinations of visual acuity and disease statistics, we illustrate in this work due to space limitations the statistics of the aforementioned three diseases, and a disease statistic under the influence of a second disease.
%
As statistical tests, we employ two-sample Student's t-tests. Furthermore, we quantify the strength of the effect~-- the increase or decrease in visual acuity~-- via the standard Cohen's d metric \cite{Cohen1988}. Cohen's d measures it in standard units, where $0.2$ stands for a small, $0.5$ for a medium, and $\ge 0.8$ for a large effect size \cite{Cohen1988}. It is calculate via equation~\ref{equation:Cohens_d}.

\begin{equation}
    d = \frac{\bar{x_1} - \bar{x_2}}{s}
%
    \label{equation:Cohens_d}
\end{equation}

Where $\bar{x_1}$ and $\bar{x_2}$ are the means of the two data sets (patient populations) and $s$ is the standard deviation for the data.

\subsection{Patient progression modeling}
\label{section:patient_progression_modeling}

For the OCT biomarker classification and patient progression modeling via visual acuity prediction, a set of prominent as well as conventional approaches from machine learning and deep learning are adapted. The machine learning library TensorFlow with Keras as its front end is utilized, while furthermore, scikit-learn is deployed for all machine learning based models and evaluation procedures. As the extracted medical data from the electronic health records of our patients originates from the documentations of ophthalmologists, these data will be defined as our ground truth. Data are extracted from the documentations as described in section \ref{section:text_processing_and_ontology} and will therefore serve as our training and test data set.

For this purpose, our patient progression visualization and modeling dashboard depicted within Fig.~\ref{figure:dashboard} is utilized to enable the annotation of the expected course of the visual acuity on site by ophthalmologists. These annotations are in turn used to obtain a comparison for the prediction capabilities based on our combined data corpus.

\subsubsection{Classification definitions and terminology}
\label{section:terminology}

For evaluation, we deploy the macro average F1-score. The macro average F1-score is calculated via the class-wise F1-scores $f$, where $F$ denotes the set of all class-wise F1-scores following our WSL classification scheme: $\textit{macro average F1-score} = \sfrac{1}{|F|} \cdot \sum_{f \in F} f$, whereas the class-wise F1-scores are in turn calculated via $\textit{F1-score} = 2 \cdot \sfrac{(\textit{precision} \cdot \textit{recall})}{(\textit{precision} + \textit{recall})}$ with $\textit{precision} = \sfrac{\textit{TP}}{\textit{TP} + \textit{FP}}$ and $\textit{recall} = \sfrac{\textit{TP}}{\textit{TP} + \textit{FN}}$ (\textit{TP} = classification true positives, \textit{FP} = false positives, and \textit{FN} = false negatives).

\subsubsection{Training and test setup}
\label{section:training_and_test_setup}

Our training setup for visual acuity prediction furthermore includes: the Glorot initializer \cite{Glorot2010} for weight initialization with the Adam optimizer as well as a batch size of the same parameterization. Over all experiments, we conducted our test runs with a randomized data set split ratio of 80/10/10 for training, validation, and test set.

\subsubsection{OCT biomarker classification}
\label{section:patient_progression_modeling_oct}

Within our proposed approach, OCT biomarkers are firstly classified using the provided OCT B-scan images (local, slice-wise classification). These B-scan images are slices through the three-dimensional, scanned back of the eye produced by the OCT scan. The local, slice-wise classifications are then combined in order to classify the whole OCT scan (global, scan-wise classification). To complete incomplete OCT biomarker documentations, our multistage system consists of an ensemble of different models for the local, slice-wise OCT classification, which in turn therefore enables the global, scan-wise OCT biomarker classification. 
For the scan-wise OCT biomarker classification, the beforehand obtained slice-wise classifications are combined via a random forest classifier as based on our classification scheme. For this combination purpose, the slice-wise obtained classification confidences are utilized and fused to a scan-wise, global class. 

Our 6 OCT biomarkers of interest are separated into two states, physiological and pathological, defining two distinct classes. These take the values interrupted (pathological) vs. preserved (physiological) for external limiting membrane (ELM) and ellipsoid zone (ellipsoid), as well as present (pathological) vs. not present (physiological) for foveal depression, retinal pigment epithelium (RPE), scars, and subretinal fibrosis, respectively. An OCT biomarker data set overview with the available OCT B-scans per OCT biomarker and related classes is shown in Table~\ref{table:OCT_biomarker_dataset_overview}, resulting in a total of 12 data subsets, for which a classification into pathological vs. physiological OCT scans is conduced. For our OCT biomarker OCT slice extraction, we determined an intermediate subset of slices with a slice range of 8 to 18 from 25 in total (8..18 out of 1..25). This range has been selected as, in our experience, most information is present within this slice range. Since different OCT scans may possess different original image resolutions, their image resolution was scaled to an initial size of $256 \times 256$ pixels for ML models and DNNs.

Based on the obtained OCT biomarker classifications, the subsequent VA modeling is realized as a time series prediction using, among others, different ML- and DL-based models such as multilayer perceptrons (MLP) \cite{Hinton1990} and recurrent neural networks \cite{Hochreiter1997, Chung2014}, also shown in Table~\ref{table:VA_modeling_results}.
%
For our classical multilayer perceptron classifier as baseline model, the following parameters were chosen as its configuration: one input, one hidden, and one output layer with a hidden layer size of 100. We utilize the ReLU activation function \cite{fukushima1975cognitron, nair2010rectified}, the Adam optimizer \cite{Kingma2015} with a standard learning rate of $0.001$ and exponential decay rates of $0.9$ and $0.999$, as well as a batch size of $32$. For all remaining parameters, standard values as provided by scikit-learn are applied.

\begin{table}[tbp]
    \centering

    \begin{tabular}{M{8cm}M{8cm}}
        \newsavebox{\bsfi}\begin{lrbox}{\bsfi}\begin{lstlisting}[linewidth=15.5cm, language=Python, basicstyle=\scriptsize, breaklines=true, frame=single, numbers=left, tabsize=1, gobble=3]
            date_{j},            date_{j+1},            date_{j+2},            ..., date_{i}
            age_{j},             age_{j+1},             age_{j+2},             ..., age_{i}
            VA_{j},              VA_{j+1},              VA_{j+2},              ..., VA_{i}
            treatment_{j},       treatment_{j+1},       treatment_{j+2},       ..., treatment_{i}
            OCT biomarker_{j,1}, OCT biomarker_{j+1,1}, OCT biomarker_{j+2,1}, ..., OCT biomarker_{i,1}
            OCT biomarker_{j,2}, OCT biomarker_{j+1,2}, OCT biomarker_{j+2,2}, ..., OCT biomarker_{i,2}
            OCT biomarker_{j,3}, OCT biomarker_{j+1,3}, OCT biomarker_{j+2,3}, ..., OCT biomarker_{i,3}
            ...,                 ...,                   ...,                   ..., ...
            OCT biomarker_{j,n}, OCT biomarker_{j+1,n}, OCT biomarker_{j+2,n}, ..., OCT biomarker_{i,n}
            additional data_{j}, additional data_{j+1}, additional data_{j+2}, ..., additional data_{i}
        \end{lstlisting}\end{lrbox}
        \subfloat[Data organization. The x-axis denotes the time (examination date) and the y-axis the data entries.]{\usebox{\bsfi}} & \\
        \subfloat[Data explanations. ]{
            \begin{tabular}{|M{1cm}|M{7cm}|}
                \hline
                Entry & Explanation \\
                \hline
                \noalign{\vskip 2pt}

                \hline
                1. & Current date in days since the patient's birth \\
                \hline
                2. & The patient's year of birth \\
                \hline
                3. & The patient's sex \\
                \hline
                4. & VA (l/r) \\
                \hline
                \noalign{\vskip 2pt}

                \hline
                \multicolumn{2}{|c|}{Treatments} \\

                \hline
                5. & Medication (l/r): Eylea or Lucentis \\
                \hline
                6.--8. & Related information: apoplexy, blood thinning, and myocardial infarction \\
                \hline
                \noalign{\vskip 2pt}

                \hline
                \multicolumn{2}{|c|}{OCT biomarkers} \\

                \hline
                9.--14. & OCT biomarkers (l/r): ELM, ellipsoid, foveal depression, RPE, scars, and subretinal fibrosis \\
                \hline
                15.--17. & Related information (l/r): central retinal thickness as well as intraretinal and subretinal fluid \\
                \hline
                \noalign{\vskip 2pt}

                \hline
                \multicolumn{2}{|c|}{Additional data} \\

                \hline
                18.--24. & Diseases (l/r): AMD, DME, and RVO as well as cataract, diabetic retinopathy, ERM, and pseudophakia \\
                \hline
            \end{tabular}} & 
        \hspace{0.2cm}
        \newsavebox{\bsfii}\begin{lrbox}{\bsfii}\begin{lstlisting}[linewidth=6cm, language=Python, basicstyle=\scriptsize, breaklines=true, frame=single, numbers=left, tabsize=1, gobble=3]
            20177, 20239, 20396, 20519
            1965, 1965, 1965, 1965
            0, 0, 0, 0
            0.8, 0.5, 0.5, 0.5
            -1,  0,  0,  0
            -1, -1, -1, -1
            -1, -1, -1, -1
            -1, -1, -1, -1
            -1, -1, -1, -1
            -1, -1, -1, -1
             1,  0, -1, -1
            -1, -1, -1, -1
            -1, -1, -1, -1
            -1, -1, -1, -1
            306, 437, -1, -1
             0, -1, -1, -1
             0, -1, -1, -1
            -1, -1, -1, -1
             1, -1, -1, -1
             2, -1, -1, -1
            -1, -1, -1, -1
            -1, -1, -1, -1
            -1, -1, -1, -1
            -1, -1, -1, -1
        \end{lstlisting}\end{lrbox}
        \subfloat[Data vectors. The x-axis denotes the time and the y-axis the data entries.]{\usebox{\bsfii}} \\
    \end{tabular}

    \caption{Ophthalmic data set overview with utilized data organization for VA prediction. To predict the VA at date $i$, a window of maximum window size 10 is utilized where $j$ is the first date of the current window ($j - i - 1$). \textbf{(a)} Essential design: For this purpose, the shown data organization from lines 1 to 10 is deployed, whereas the data within the fields ``treatment'', ``OCT biomarkers'', and ``additional data'' cover multiple entries as detailed in (b). \textbf{(b)} Detailed data organization, outlining the fields from (a). Here, ``l/r'' denotes the availability of data for both eyes (left/right). In Fig.~\ref{figure:dashboard}, two entries for RPE exist, RPE (availability) and RPE detachment (state, physiological or pathological), which have been combined into one entry. \textbf{(c)} When translating the information from text strings in (a) and (b) via numerical translation to values in (c), value-based data vectors are obtained that are applied to the models' learning and evaluation procedures. In (c), ``$-1$'' denotes a numeric placeholder when no information is present within the respective data fields. For the sake of simplicity, the processing step of numerical translation was omitted. For example, the data entry of a patient's sex can result in three different values: $0$ for male, $1$ for female, and $2$ for patients of diverse sex. Shown are the values for the first window of size 4 in Fig.~\ref{figure:dashboard}.}
    \label{table:VA_modeling_dataset_overview}
\end{table}

\subsubsection{Visual acuity prediction}
\label{section:patient_progression_modeling_va}

To allow a WSL-based grouping of VA values, the logMAR score of each VA value is derived via $\textit{VA}_\textit{logMAR} = -\log_{10}\textit{VA}_\textit{dec}$ \cite{Schulze2006}. We define a decimal VA range of $0.04$ to $2.0$, corresponding to a logMAR range of $1.4$ to $-0.3$. The visual acuity delta ($\Delta\textit{VA}_i$) of the examination $i$ is calculated by comparing two adjacent logMAR VA values via $\Delta\textit{VA}_i = \textit{VA}_i - \textit{VA}_{i - 1}$. A threshold of $0.1$ logMAR units divides examination $i$ depending on its progression: $\Delta\textit{VA}_i < -0.1$ is considered a progression winner, $-0.1 \leq \Delta\textit{VA}_i \leq 0.1$ a progression stabilizer, and $\Delta\textit{VA}_i > 0.1$ a progression losers (see also Fig.~\ref{figure:VA_modeling}). The threshold of $0.1$ logMAR units has been selected in order to to enable a reliable categorization of progression winners, stabilizers, and losers where an improvement, stabilization, or aggravation in visual acuity is apparent.

Table~\ref{table:VA_modeling_dataset_overview} gives an overview of our ophthalmic data set with its data organization for VA prediction as it is provided to our predictive models. 
In Table~\ref{table:VA_modeling_dataset_overview}, the data organization of an exemplary time window of 4 VA measurements is shown in \textbf{(a)} for the first 10 of 24 lines of the feature data vector, while \textbf{(b)} explains in details all 24 lines of medical features associated with each VA's feature vector, including ``treatment'', ``OCT biomarker'', as well as ``additional data''. The full data feature vector is then translated to numerical values and feed to the ML/DL models, illustrated with an example in \textbf{(c)}. In \textbf{(c)}, ``$-1$'' denotes the numeric placeholder when no information is present within the respective data fields. To predict the VA at a given date, machine learning models typically require a fixed data input size, i.e., a matrix or vector of fixed dimensions. In \textbf{(a)}, shown are the values for the first time window of size 4 in Fig.~\ref{figure:dashboard}. Our analyses currently include only IVOM therapies, while within our data vectors, no information on therapeutic interventions such as operations is present currently.

In order to make predictions for time windows of different sizes, we define a matrix with predefined dimensions of 24 rows (see medical feature vector in Table~\ref{table:VA_modeling_dataset_overview}, \textbf{(b)}) and 10 columns, which corresponds to 10 time steps or date entries with available information. The minimum time window size is 4, which means that the first 4 of in total 10 columns of the matrix are filled with patient data as exemplary shown in Table~\ref{table:VA_modeling_dataset_overview}, \textbf{(b)}. The remaining 6 columns are set to ``$-1$'', especially when no more temporal information is available. The window is~-- as far as more temporal data are available~-- iteratively increased by one, i.e., in each iteration, one column more is filled with values. The subsequent VA is modeled, for which it is determined whether the VA improves, remains constant, or deteriorates (WSL classification scheme). However, at maximum, 10 time steps (columns) are used given that the visual acuity for the 11th time step is known. Formally, all models in Table~\ref{table:VA_modeling_results} require a vector as input. Thus, the matrix was reshaped into a vector of size $1 \times 240$ to be used in model training and testing. The MLP model described in section~\ref{section:patient_progression_modeling_oct} with its configuration is again utilized as baseline model. Given our MLP model for visual acuity prediction, a meta model is realized that classifies the predicted VA values via our WSL classification scheme. Since the MLP predicts the VA values, our so-called MLP-LDA model utilizes these predictions by further classifying them via a linear discriminant analysis (LDA) \cite{Hastie2009} into our WSL-based classes. To this end, the shown data vectors in Table~\ref{table:VA_modeling_dataset_overview}c are extended with the visual acuity predictions of our MLP model in each time step. MLP-LDA processes these extended data vectors while correcting the previously obtained visual acuity prediction of MLP.

For our ophthalmologists' study, two different sets of ophthalmologist(s) were recruited for a first evaluation study: our main ophthalmologist (ophthalmologist I) and eight other ophthalmologists to further validate the results of ophthalmologist I (ophthalmologist set II). Our motivation was to provide the main ophthalmologist with the full test data set, while group II received randomized subsets of our test set. We also aimed at choosing multiple participants, yet time and effort were limiting factors. All participants were selected from the Department of Ophthalmology at the Klinikum Chemnitz gGmbH in Chemnitz, Germany and they have all subspecialty training in eye diseases. Training of our ophthalmologist I is fellowship-level. The participant group II has a span of training levels (four fellow level, three specialist, and one senior specialist for retinal surgery). Their experience is assumed based on their respective training levels. The main ophthalmologist is author of this study, he has thus potentially a notably higher experience in the field of AMD/DME/RVO as his training level otherwise implies, and the senior specialist is author as well. Within our experiment, the ophthalmologists utilized our dashboard for their predictions (Fig.~\ref{figure:dashboard}), for which the task was to predict the visual acuity value for the next point in time. The test data set for ophthalmologist I contained 1494 samples, while the eight additional doctors received randomized subsets of the test set of ophthalmologist I, whereby each subset contained between 50--100 samples. Participants were presented with the following medical information, which is also visible from our dashboard: the past visual acuity values, central retinal thickness, OCT biomarker documentation, diagnoses (AMD, DME, RVO, and also others), general information (gender and age), IVOMs, and additional medical data. The AI models were provided with precisely the same information (cf. Table~\ref{table:VA_modeling_dataset_overview}).

\section{Test results, evaluation, and discussion}
\label{section:evaluation}

The following sections provide first our evaluation regarding the predictive statistics of therapy winners, stabilizers, and losers (section~\ref{section:predictive_statistics}). To allow the inclusion of OCT biomarkers into the patient progression and VA modeling process, incomplete or missing OCT documentations are completed (section~\ref{section:oct_biomarker_classification}). Given each patient's medical data, the following patient progression and VA prediction utilizes the resulting OCT biomarker completions (section~\ref{section:visual_acuity_prediction}).

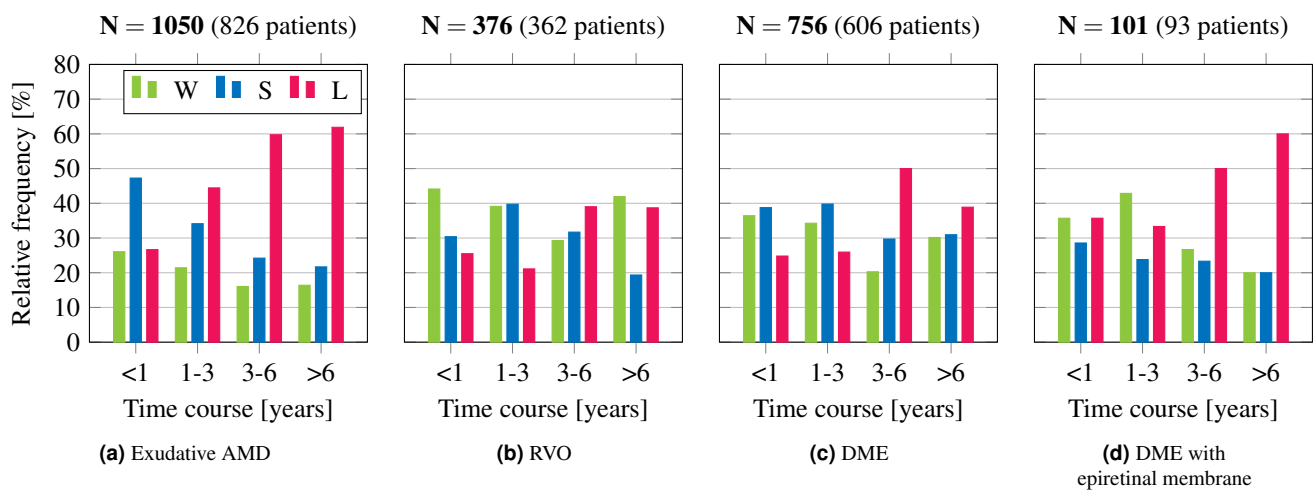
\begin{figure}[tb]
    \definecolor{LimeGreen}{HTML}{8DC73E}
    \definecolor{Red}{HTML}{ED135A}

    \centering

    \subfloat[Exudative AMD]{
        \begin{tikzpicture}
            \begin{axis}[
             ybar, bar width=0.15cm, enlarge x limits=0.25, symbolic x coords={<1, 1-3, 3-6, >6},
             width=0.3\textwidth, height=0.3\textwidth,
             title={$\mathbf{N = 1050}$ (826 patients)}, xlabel={Time course [years]}, ylabel={Relative frequency [\%]},
             ymin=0.0, ymax=0.8,
             ytick={0.0, 0.1, 0.2, 0.3, 0.4, 0.5, 0.6, 0.7, 0.8, 0.9}, yticklabels={0, 10, 20, 30, 40, 50, 60, 70, 80, 90},
             xtick=data, ymajorgrids=true,
             legend columns=-1, legend style={column sep=1ex}]
                \addplot[draw=LimeGreen,   fill=LimeGreen]   coordinates {(<1, 0.2606) (1-3, 0.2143) (3-6, 0.1604) (>6, 0.1640)};
                \addplot[draw=custom_blue, fill=custom_blue] coordinates {(<1, 0.4727) (1-3, 0.3413) (3-6, 0.2421) (>6, 0.2169)};
                \addplot[draw=Red,         fill=Red]         coordinates {(<1, 0.2666) (1-3, 0.4444) (3-6, 0.5975) (>6, 0.6190)};

                \legend{W, S, L}
            \end{axis}
        \end{tikzpicture}
        \label{figure:wsl_distribution_a}}
    \hfill
    \subfloat[RVO]{
        \begin{tikzpicture}
            \begin{axis}[
             ybar, bar width=0.15cm, enlarge x limits=0.25, symbolic x coords={<1, 1-3, 3-6, >6},
             width=0.3\textwidth, height=0.3\textwidth,
             title={$\mathbf{N = 376}$ (362 patients)}, xlabel={Time course [years]}, ylabel={},
             ymin=0.0, ymax=0.8,
             ytick={0.0, 0.1, 0.2, 0.3, 0.4, 0.5, 0.6, 0.7, 0.8, 0.9}, yticklabels={},
             xtick=data, ymajorgrids=true]
                \addplot[draw=LimeGreen,   fill=LimeGreen]   coordinates {(<1, 0.4412) (1-3, 0.3913) (3-6, 0.2927) (>6, 0.4194)};
                \addplot[draw=custom_blue, fill=custom_blue] coordinates {(<1, 0.3039) (1-3, 0.3975) (3-6, 0.3170) (>6, 0.1935)};
                \addplot[draw=Red,         fill=Red]         coordinates {(<1, 0.2549) (1-3, 0.2111) (3-6, 0.3902) (>6, 0.3870)};
            \end{axis}
        \end{tikzpicture}
        \label{figure:wsl_distribution_b}}
    \hfill
    \subfloat[DME]{
        \begin{tikzpicture}
            \begin{axis}[
             ybar, bar width=0.15cm, enlarge x limits=0.25, symbolic x coords={<1, 1-3, 3-6, >6},
             width=0.3\textwidth, height=0.3\textwidth,
             title={$\mathbf{N = 756}$ (606 patients)}, xlabel={Time course [years]}, ylabel={},
             ymin=0.0, ymax=0.8,
             ytick={0.0, 0.1, 0.2, 0.3, 0.4, 0.5, 0.6, 0.7, 0.8, 0.9}, yticklabels={},
             xtick=data, ymajorgrids=true]
                \addplot[draw=LimeGreen,   fill=LimeGreen]   coordinates {(<1, 0.3643) (1-3, 0.3426) (3-6, 0.2028) (>6, 0.3016)};
                \addplot[draw=custom_blue, fill=custom_blue] coordinates {(<1, 0.3876) (1-3, 0.3979) (3-6, 0.2972) (>6, 0.3095)};
                \addplot[draw=Red,         fill=Red]         coordinates {(<1, 0.2481) (1-3, 0.2595) (3-6, 0.5000) (>6, 0.3889)};
            \end{axis}
        \end{tikzpicture}
        \label{figure:wsl_distribution_c}}
    \hfill
    \subfloat[DME with \\ epiretinal membrane]{
        \begin{tikzpicture}
            \begin{axis}[
             ybar, bar width=0.15cm, enlarge x limits=0.25, symbolic x coords={<1, 1-3, 3-6, >6},
             width=0.3\textwidth, height=0.3\textwidth,
             title={$\mathbf{N = 101}$ (93 patients)}, xlabel={Time course [years]}, ylabel={},
             ymin=0.0, ymax=0.8,
             ytick={0.0, 0.1, 0.2, 0.3, 0.4, 0.5, 0.6, 0.7, 0.8, 0.9}, yticklabels={},
             xtick=data, ymajorgrids=true]
                \addplot[draw=LimeGreen,   fill=LimeGreen]   coordinates {(<1, 0.3571) (1-3, 0.4286) (3-6, 0.2667) (>6, 0.2000)};
                \addplot[draw=custom_blue, fill=custom_blue] coordinates {(<1, 0.2857) (1-3, 0.2380) (3-6, 0.2333) (>6, 0.2000)};
                \addplot[draw=Red,         fill=Red]         coordinates {(<1, 0.3571) (1-3, 0.3333) (3-6, 0.5000) (>6, 0.6000)};
            \end{axis}
        \end{tikzpicture}
        \label{figure:wsl_distribution_d}}

    \caption{\textbf{(a--c)} Real-world winner, stabilizer, and loser distribution (WSL classification scheme) for exudative AMD, RVO, and DME. \textbf{(d)} Distribution for the disease DME under exemplary medical co-factor of an epiretinal membrane. The shown results are based on our disease statistics (Fig.~\ref{figure:workflow}), whereas $\mathbf{N}$ denotes the number of eyes for the given number of patients.}
    \label{figure:wsl_distribution}
\end{figure}

\begin{table}[tb]
    \newcommand{\hpleq}{\hphantom{<{}}}
    \centering

    \begin{tabular}{|P{5.5cm}|P{2cm}|P{2cm}|P{2cm}|P{2cm}|P{2cm}|}
        \hline
        Condition & Significant? & t & p value & Cohen's d \\
        \hline
        \noalign{\vskip 2pt}

        \hline
        AMD, <1y compared to 1-3y    & 1 & $3.09$ & $\hpleq 0.0020$ & $0.30$ \\
        \hline
        AMD, <1y compared to >3y     & 1 & $6.49$ & $<0.0001$ & $\mathbf{0.70}$ \\
        \hline
        RVO, <1y compared to >3y     & 0 & $1.89$ & $\hpleq 0.0607$ & $0.38$ \\
        \hline
        DME, <1y compared to >3y     & 1 & $3.19$ & $\hpleq 0.0016$ & $0.40$ \\
        \hline
        DME compared to DME+ERM, >6y & 0 & $1.16$ & $\hpleq 0.2477$ & $0.32$ \\
        \hline
    \end{tabular}

    \caption{Summary of the statistical test results regarding the significant deterioration of the visual acuity. The deterioration is expressed by an increase in the loser fraction. For each condition tested, t-tests were conducted using an alpha parameter of $0.05$ (false discovery rate). For example, the first row compares the amount of losers in under 1 year time passed for the disease AMD with the condition of 1-3 years passed. The values refer to the t, p, and Cohen's d value. Cohen's d measures the magnitude of the effect (effect size). A value of 0.2 stands for a small, 0.5 for a medium and $\ge 0.8$ for a large effect, whereby medium and large effects are marked in bold.}
    \label{tab:stat_test_results}
\end{table}

\subsection{Predictive statistics}
\label{section:predictive_statistics}

For the following predictive statistics, a statistical analysis was conducted for our diseases exudative AMD, RVO, and DME. The progression of the VA was classified into therapy winners, stabilizers, and losers (WSL classification scheme) based on the first and the last VA measurement of each patient (section~\ref{section:descriptive_statistics}). The size of our data corpus and its harmonization as described in sections~\ref{section:system_architecture} to \ref{section:data_fusion_and_cleaning} allows different kinds of statistical surveys, e.g., separated according to disease, time periods, and comorbidities. The data originate from a large- and daily-operating medical hospital (German hospital of maximum care level) and thus indicate effects of real-world scenarios.

Our statistical investigations consequently allow us to make statistical predictions under real-life conditions for questions such as ``If a patient has exudative AMD, what are the future prognoses for this patient?''. For the three aforementioned diseases, the outcomes are shown in Fig.~\ref{figure:wsl_distribution}. Especially for AMD, at average, a deterioration of the visual acuity over time is observable.
A more fine-grained analysis reveals effects over time since we split the data of a disease into patients with short and longer progression. The time course refers to the time difference (in years) between the first and the last VA measurement of individual patients. The total data of $\mathbf{N} = \text{1050}$ eyes of AMD is now divided into 4 substatistics with different time windows, whereby, for example, the data in the first time bin of under 1 year is about $25~\%$. We found, with regard to longer disease time courses, the proportion of losers increases further till $\ge 60$~\% for time courses of >6 years and longer. Note that our WSL group definition using $\Delta\textit{VA}_\textit{logMAR}$ thresholds of $0.1$ is fixed for all time windows, which might be regarded as a somewhat harsh criterion for long time scales.
We perform two-sample Student's t-tests to analyze the statistical significance of the deterioration of the visual acuity, shown in Table~\ref{tab:stat_test_results}. To avoid thresholding effects, we perform the tests directly at the raw delta logMAR values. We found a strong significant effect for the disease AMD ($p \leq 0.0001$), no significance for RVO ($p = 0.0607$), and a weak significant effect for DME ($p = 0.0016$). A full combination of all statistical tests can be found in Table~\ref{tab:stat_test_results_extensive}. In addition, we employed the Cohen's d measurement that shows the normalized strength of the effect, i.e., the amount of the increase in the therapy loser fraction (section~\ref{section:descriptive_statistics}). We observed a deterioration of the visual acuity in AMD with a large/medium effect (Table~\ref{tab:stat_test_results}), while the other diseases arouse smaller effects. This means that more and more patients are experiencing deterioration of vision over longer periods of time, especially for AMD.

The representation of our diseases in combination with medical co-factors (comorbidities) is shown in Fig.~\ref{figure:wsl_distribution_d} and can be performed as a proof of concept. It illustrates the influence of an epiretinal membrane (ERM) on the disease DME. Yet, if DME and epiretinal membrane occur simultaneously, it becomes apparent that only about 12--25 patients are included in each substatistic and the direct comparison with the DME-only group would not yet stand up to statistical tests (Table~\ref{tab:stat_test_results}). For such substatistics, more data will be needed in the future, e.g., by merging several ophthalmic hospitals into a common research data infrastructure.

\begin{table}[tb]
    \centering

    \begin{tabular}{|M{3cm}||M{1.8cm}|M{1.8cm}|M{1.8cm}|M{1.8cm}|M{1.8cm}||M{2.4cm}|}
        \cline{2-7}
        \multicolumn{1}{c|}{} & \multicolumn{5}{c||}{Slice-wise F1-score [\%]} & Scan-wise F1-score [\%] \\
        \cline{2-7}
        \noalign{\vskip 2pt}

        \hline
        OCT biomarker & Logistic Regression \cite{Hastie2009} & Random Forest Classifier \cite{Hastie2009} & Multilayer Perceptron Classifier \cite{Hinton1990} & DenseNet\-201 \cite{Huang2017} & ResNet\-152V2 \cite{He2016a} & Random Forest Classifier \cite{Hastie2009} \\
        \hline
        \noalign{\vskip 2pt}

        \hline
        ELM                 & $85.4 \pm 2.1$ & $92.1 \pm 1.4$ & $95.3 \pm 0.9$ & $96.2 \pm 0.6$ & $\mathbf{98.3 \pm 0.5}$ & $\mathbf{99.9 \pm 0.1}$ \\
        \hline
        Ellipsoid           & $82.7 \pm 1.7$ & $93.8 \pm 1.4$ & $96.1 \pm 0.7$ & $97.3 \pm 1.2$ & $\mathbf{98.7 \pm 1.1}$ & $\mathbf{99.9 \pm 0.1}$ \\
        \hline
        Foveal depression   & $56.0 \pm 3.4$ & $69.4 \pm 0.9$ & $77.3 \pm 2.5$ & $75.9 \pm 2.3$ & $\mathbf{77.5 \pm 2.4}$ & $\mathbf{99.9 \pm 0.1}$ \\
        \hline
        RPE                 & $50.4 \pm 1.4$ & $63.6 \pm 0.9$ & $64.0 \pm 1.1$ & $\mathbf{69.0 \pm 1.0}$ & $67.3 \pm 0.9$ & $\mathbf{94.8 \pm 0.5}$ \\
        \hline
        Scars               & $63.8 \pm 0.9$ & $72.2 \pm 1.0$ & $74.5 \pm 1.3$ & $77.2 \pm 1.2$ & $\mathbf{77.6 \pm 1.0}$ & $\mathbf{98.3 \pm 1.2}$ \\
        \hline
        Subretinal fibrosis & $69.3 \pm 1.5$ & $70.6 \pm 0.6$ & $74.5 \pm 0.4$ & $\mathbf{75.1 \pm 0.2}$ & $74.6 \pm 0.3$ & $\mathbf{96.2 \pm 0.9}$ \\
        \hline
        \noalign{\vskip 2pt}

        \hline
        Mean & $67.9$ & $77.0$ & $80.3$ & $81.8$ & $\mathbf{82.3}$ & $\mathbf{98.2}$ \\
        \hline
    \end{tabular}

    \caption{OCT biomarker classification results for the slice-wise as well as the scan-wise OCT biomarker classification. We conducted our test runs with a randomized data set split ratio of 80/10/10 for training, validation, and test set averaged over five runs. For our OCT biomarker data set overview, see Table~\ref{table:OCT_biomarker_dataset_overview}.}
    \label{table:OCT_biomarker_classification_results}
\end{table}

\begin{table}[tb]
    \renewcommand{\arraystretch}{1.1}
    \centering

    \begin{tabular}{|M{8cm}|M{4cm}|M{4cm}|}
        \cline{2-3}
        \multicolumn{1}{c|}{} & \multicolumn{2}{c|}{Macro average F1-score [\%]} \\
        \cline{2-3}
        \noalign{\vskip 2pt}

        \hline
        Model & Feature vectors containing VA values only & Feature vectors containing additional medical data \\
        \hline
        \noalign{\vskip 2pt}

        \hline
        \multicolumn{3}{|c|}{Model predictions} \\
        \hline
        \noalign{\vskip 2pt}

        \hline
        Statistical estimator                              & $32.8 \pm 0.4$ & / \\
        \hline
        MA estimator                                       & $29.6 \pm 0.3$ & / \\
        \hline
        Weighted MA estimator                              & $22.6 \pm 0.4$ & / \\
        \hline
        Bagging regressor \cite{Hastie2009}                & $\mathbf{40.8 \pm 0.2}$ & $\mathbf{41.3 \pm 0.5}$ \\
        \hline
        Random forest regressor \cite{Hastie2009}          & $39.9 \pm 0.1$ & $40.7 \pm 0.4$ \\
        \hline
        RNN with LSTM \cite{Rumelhart1986, Hochreiter1997} & $37.6 \pm 0.3$ & $39.0 \pm 0.3$ \\
        \hline
        RNN with GRU \cite{Rumelhart1986, Chung2014}       & $38.1 \pm 0.3$ & $39.1 \pm 0.1$ \\
        \hline
        MLP \cite{Hinton1990}                              & $\mathbf{40.2 \pm 0.4}$ & $\mathbf{44.6 \pm 0.6}$ \\
        \hline
        MLP-LDA \cite{Hinton1990, Hastie2009}              & $\mathbf{44.9 \pm 4.5}$ & $\mathbf{69.0 \pm 5.2}$ \\
        \hline
        \noalign{\vskip 4pt}

        \hline
        \multicolumn{3}{|c|}{Model predictions: control experiments} \\
        \hline
        \noalign{\vskip 2pt}

        \hline
        MLP without annotated OCT biomarkers & / & $42.9 \pm 0.4$ \\
        \hline
        MLP without classified OCT biomarkers & / & $42.7 \pm 0.6$ \\
        \hline
        MLP-LDA without annotated OCT biomarkers & / & $63.1 \pm 4.7$ \\
        \hline
        MLP-LDA without classified OCT biomarkers & / & $62.8 \pm 3.1$ \\
        \hline
        \noalign{\vskip 4pt}

        \hline
        \multicolumn{3}{|c|}{Annotations} \\
        \hline
        \noalign{\vskip 2pt}

        \hline
        Ophthalmologists & \multicolumn{2}{c|}{$\mathbf{57.8}$ and $\mathbf{50.0 \pm 10.7}$} \\
        \hline
    \end{tabular}

    \caption{Visual acuity modeling results overview of our approaches with feature vectors containing visual acuity values only / additional medical data in the form of our completed OCT biomarker documentations (Table~\ref{table:VA_modeling_dataset_overview}b), and our annotations (ophthalmologists). All results are averaged over five runs. ``/'' denotes model predictions without additional medical data and VA values only. The three best model prediction results are highlighted each in bold. Note, we reported the results for both sets of ophthalmologists. For our VA modeling data set overview with data explanations and vectors, see Table~\ref{table:VA_modeling_dataset_overview}. Figures~\ref{figure:VA_modeling_results_plot} and \ref{figure:VA_modeling_results_CM} as well as Table~\ref{table:VA_modeling_results_extensive} show our evaluation results in further detail.}
    \label{table:VA_modeling_results}
\end{table}

\subsection{OCT biomarker classification}
\label{section:oct_biomarker_classification}

Table~\ref{table:OCT_biomarker_classification_results} shows our classification results for the slice-wise and the scan-wise OCT classification using different prominent approaches from machine learning and deep learning averaged over five runs. In comparison, the selected models DenseNet-201 \cite{Huang2017} and ResNet-152V2 \cite{He2016a} show the best classification accuracies in F1-score with mean classification accuracies of $81.8$ and $82.3$~\% for our 6 OCT biomarkers. While the biomarkers RPE and subretinal fibrosis are best classified with the network DenseNet-201, the best results for the four other biomarkers ELM, ellipsoid, foveal depression, and scars are shown by the ResNet-152V2 network. When comparing the results for the different OCT biomarkers, ELM and ellipsoid show the best accuracies to classifying them given the OCT slices, whereas RPE and subretinal fibrosis represent the more challenging OCT biomarkers with reduced accuracies. For the following scan-wise OCT classification, a random forest classifier \cite{Hastie2009} was sufficient in order to achieve high classification accuracies over all biomarkers with single scores of up to $99.9$~\%. These scores were obtained for the biomarkers ELM, ellipsoid, and foveal depression. Finally, we obtain the best resulting mean classification accuracies in F1-score over all OCT biomarkers of $82.3$ (slice-wise, ResNet-152V2), and of $98.2$~\% (scan-wise, random forest classifier).

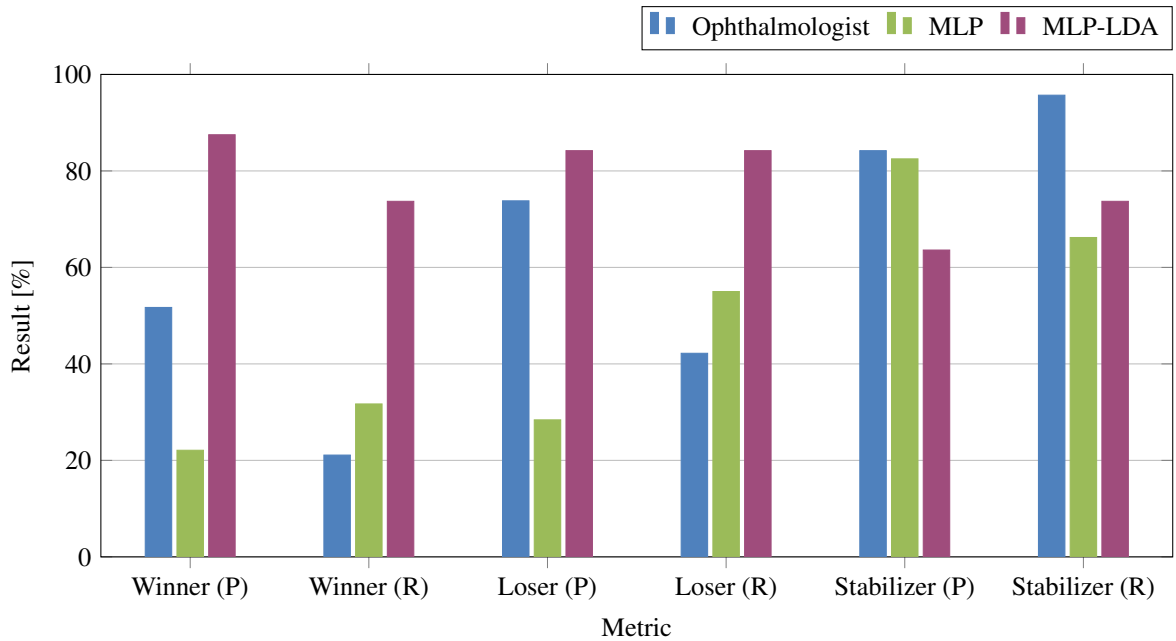
\begin{figure}[tb]
    \definecolor{bblue}{HTML}{4F81BD}
    \definecolor{ggreen}{HTML}{9BBB59}
    \definecolor{ppurple}{HTML}{9F4C7C}

    \centering

    \begin{tikzpicture}
        \begin{axis}[
         ybar, enlarge x limits=0.1,
         symbolic x coords={Winner (P), Winner (R), Loser (P), Loser (R), Stabilizer (P), Stabilizer (R)},
         width=0.9\textwidth, height=8cm,
         xlabel={Metric}, ylabel={Result [\%]},
         ymin=0, ymax=100,
         xtick=data, ymajorgrids=true,
         legend columns=-1, legend style={at={(1, 1.05)}, anchor=south east, column sep=1ex}]
            \addplot[draw=bblue,   fill=bblue]   coordinates {(Winner (P), 51.7) (Winner (R), 21.1) (Loser (P), 73.8) (Loser (R), 42.2) (Stabilizer (P), 84.2) (Stabilizer (R), 95.7)};
            \addplot[draw=ggreen,  fill=ggreen]  coordinates {(Winner (P), 22.1) (Winner (R), 31.7) (Loser (P), 28.4) (Loser (R), 55.0) (Stabilizer (P), 82.5) (Stabilizer (R), 66.2)};
            \addplot[draw=ppurple, fill=ppurple] coordinates {(Winner (P), 87.5) (Winner (R), 73.7) (Loser (P), 84.2) (Loser (R), 84.2) (Stabilizer (P), 63.6) (Stabilizer (R), 73.7)};

            \legend{Ophthalmologist, MLP, MLP-LDA}
        \end{axis}
    \end{tikzpicture}

    \caption{VA modeling results' precision (P) and recall (R) plot for our annotations (main ophthalmologist) as well as our approaches to VA prediction with MLP and MLP-LDA based on our WSL classification scheme (see also Table~\ref{table:VA_modeling_results_extensive}).}
    \label{figure:VA_modeling_results_plot}
\end{figure}

\subsection{Visual acuity prediction}
\label{section:visual_acuity_prediction}

The principles of VA prediction are illustrated in Fig.~\ref{figure:VA_modeling} for an exemplary patient whose first diagnosis was cataract in both eyes and DME in the right eye in 02/2014. Shown is the VA progression over a timespan of 1 year, in which the patient had 6 IVOMs with Eylea and Lucentis. For visual acuity prediction, we consider the subsequences of measured visual acuity values with their additional data for each patient, for which the future visual acuity value is predicted. Our model predicts the $(i + 4)$th VA value from a time window of the previous four VA measurements, whereby we use a growing time window of size 4 up to a size of 10 VA measurements, e.g., the minimum interval $[i, i + 3]$ with $i = 0, 1, \cdots, i_\textit{max} - 4$. This approach has been selected to account for the different sizes of visual acuity sequences. A minimum window size is enforced in order to enable a more reliable prediction, whereby patients with an insufficient amount of measurements are not considered. Additionally, a maximum window size was defined as, in our experience, larger window sizes can lead to reduced prediction scores. The last documented VA measurement is defined to be $i_\textit{max}$. The model uses medical patient with the aforementioned growing window size, which is reformatted as data input matrix as shown in Table~\ref{table:VA_modeling_dataset_overview}b. For instance, the 5th VA value will be predicted based on the time interval provided by the 1st to the 4th VA measurement, whereas the 6th VA value will be predicted via the 1st to the 5th measurement. For evaluation, the horizontal lines indicate our thresholds for therapy winners, stabilizers, and losers (see also section~\ref{section:predictive_statistics}). Finally, a classification based on our WSL classification scheme is carried out for our predicted visual acuity values. The obtained WSL-based classification is then compared to the original classification of our ground truth.

\subsubsection{Evaluation principles}
\label{section:evaluation_principles}

Out of 49\,000 patients, 7\,878 patients with VA series of length $i_\textit{max} \geq 5$ exist within our data, resulting in over 100\,000 separate VA series of length 5. This minimum sequence length of 5 VA measurements has been selected as, in our experience, shorter sequences may not allow for a reliable data for the visual acuity prediction. With the three diseases AMD, DME, and RVO, 1\,496 patients with VA series of length $\geq$ 5 exist, resulting in 14\,026 separate VA series of length 5.
%
For evaluation, all visual acuity values with their time steps are considered. Thus, all predictions as shown in Fig.~\ref{figure:VA_modeling} are utilized, for which all time steps are evaluated regarding their resulting WSL-based classification. The VA-based prediction accuracy is calculated via all VA predictions and related local $\Delta\textit{VA}_i$. For the modeling process, the completed OCT documentations (section~\ref{section:oct_biomarker_classification}) and the related additional data are retrieved. For this purpose, Table~\ref{table:VA_modeling_dataset_overview} gives an extensive overview of the leveraged data set as well as the related data organization.

\subsubsection{Visual acuity prediction results}
\label{section:visual_acuity_prediction_results}

Table~\ref{table:VA_modeling_results} shows our prediction results using different statistical approaches as well as prominent approaches from machine learning and deep learning averaged over five runs. These approaches encompass estimators such as statistical estimators and moving average estimators (MA), regressors, recurrent neural networks, and multilayer perceptrons. Our statistical estimator predicts visual acuity progressions utilizing the statistical distribution of our WSL classification scheme within our train set. The MA estimator averages the given window of VA values, whereas the weighted MA estimator weights recent VA values more strongly. We consider the statistical estimator, MA estimator, and the weighted MA estimator in order to formulate a baseline prediction(s).

Whereas our baseline approaches, our estimators and regressors, result in prediction accuracies of up to $40.8$~\% in macro average F1-score (bagging regressor \cite{Hastie2009}), a more realistic setting includes the completed OCT documentations with the additional data shown in our data organization table such as OCT biomarkers. Our MLP-based predictor results in the second highest prediction accuracy, whereas the addition of OCT biomarkers allows their inclusion in the VA modeling process, resulting in an accuracy of $44.6$~\% with an improvement by $+4.4$~\% (Table~\ref{table:VA_modeling_results}). Therefore, the inclusion of OCT biomarkers allows their modeling as crucial information and influential visual factors when no OCT classifications are provided.

With MLP-LDA, we obtain a final prediction accuracy of $69$~\% (Table~\ref{table:VA_modeling_results}), which corresponds to an improvement by $+24.4$~\% for MLP-LDA in comparison to MLP. The ophthalmologist reaches a score of $57.8$~\%. Figure~\ref{figure:VA_modeling_results_plot} gives an overview of our obtained main results in precision and recall for therapy winners, stabilizers, and losers. Considering the class-wise scores, MLP-LDA strikes a balance between all three progression groups, whereas a trade-off between therapy winners and losers with therapy stabilizers is observable in comparison to MLP. Finally, Table~\ref{table:VA_modeling_results_extensive} shows an extensive VA modeling results overview with confusion matrices and class-wise recall and precision results of all VA modeling experiments with MLP, MLP-LDA, and the human reference annotation results from the ophthalmologist. We conclude that treatment winners and losers are predicted within the same range as the ophthalmologist in both recall and precision, which is a promising result. However, treatment stabilizers are predominantly present when observing the visual acuity values of adjacent time steps. For this reason, an improvement of the VA prediction for stabilizers has to be enforced to realize a (semi-)automated recommender system.

Finally, in order to further validate the annotations of the ophthalmologist, we evaluated the annotations of eight different additional ophthalmic doctors given randomized subsets of our test set. Table~\ref{table:VA_modeling_results_extensive_8_doctors}a shows their mean and standard deviations for precision, recall, and F1-score as well as their overall macro average F1-score. Additionally, their confusion matrices are depicted (Table~\ref{table:VA_modeling_results_extensive_8_doctors}b). We obtain a prediction accuracy in macro average F1-score of $50.0 \pm 10.7$~\%. The minimum and maximum scores are $37.7$ and $69.4$~\%, resulting in a range of $31.7$~\%. 

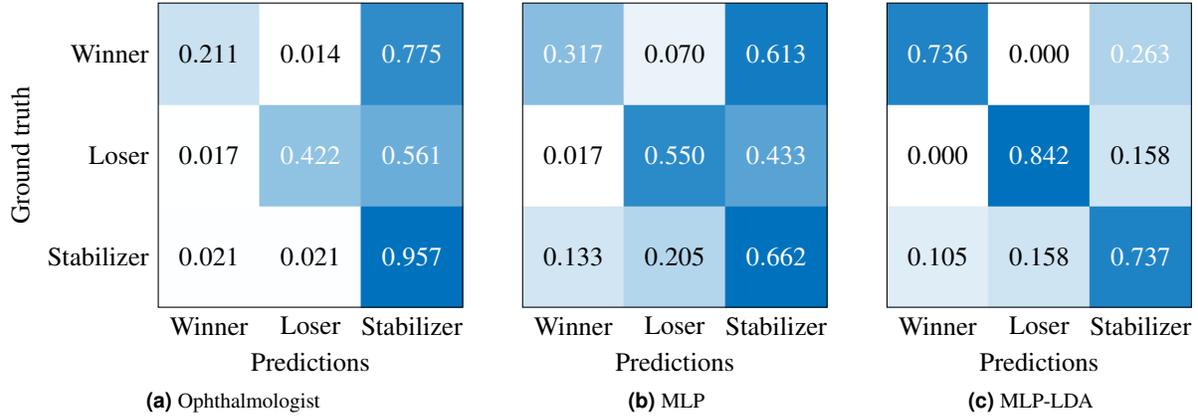
\begin{figure}[tb]
    \centering

    \subfloat[Ophthalmologist]{
        \begin{tikzpicture}
            \begin{axis}[
             xlabel={Predictions}, ylabel={Ground truth},
             xticklabels={Winner, Loser, Stabilizer}, xtick={0, 1, 2}, xtick style={draw=none},
             yticklabels={Stabilizer, Loser, Winner}, ytick={0, 1, 2}, ytick style={draw=none},
             width=0.32\textwidth, height=0.32\textwidth,
             enlargelimits=false, axis on top,
             colormap={summap}{color=(white) color=(custom_blue)},
             nodes near coords, nodes near coords align={center},
             nodes near coords black white/.style={
                 small value/.style={text=black},
                 large value/.style={text=white},
                 every node near coord/.style={
                     /pgf/number format/.cd, precision=3, fixed, fixed zerofill, /tikz/.cd,
                     check for zero/.code={
                         \pgfmathfloatifflags{\pgfplotspointmeta}{0}{\pgfkeys{/tikz/coordinate}}{
                             \begingroup
                                 \pgfkeys{/pgf/fpu}
                                 \pgfmathparse{\pgfplotspointmeta<#1}
                                 \global\let\result=\pgfmathresult
                             \endgroup
                             \pgfmathfloatcreate{1}{1.0}{0}
                             \let\ONE=\pgfmathresult
                             \ifx\result\ONE
                                 \pgfkeysalso{/pgfplots/small value}
                             \else
                                 \pgfkeysalso{/pgfplots/large value}
                             \fi
                         }
                     },
                     check for zero
                 }
             },
             nodes near coords black white=0.24]

                \addplot[matrix plot*, point meta=explicit] file
                 {data/CM_ophthalmologist.csv};
            \end{axis}
        \end{tikzpicture}}
    \quad
    \subfloat[MLP]{
        \begin{tikzpicture}
            \begin{axis}[
             xlabel={Predictions},
             xticklabels={Winner, Loser, Stabilizer}, xtick={0, 1, 2}, xtick style={draw=none},
             yticklabels={}, ytick style={draw=none},
             width=0.32\textwidth, height=0.32\textwidth,
             enlargelimits=false, axis on top,
             colormap={summap}{color=(white) color=(custom_blue)},
             nodes near coords, nodes near coords align={center},
             nodes near coords black white/.style={
                 small value/.style={text=black},
                 large value/.style={text=white},
                 every node near coord/.style={
                     /pgf/number format/.cd, precision=3, fixed, fixed zerofill, /tikz/.cd,
                     check for zero/.code={
                         \pgfmathfloatifflags{\pgfplotspointmeta}{0}{\pgfkeys{/tikz/coordinate}}{
                             \begingroup
                                 \pgfkeys{/pgf/fpu}
                                 \pgfmathparse{\pgfplotspointmeta<#1}
                                 \global\let\result=\pgfmathresult
                             \endgroup
                             \pgfmathfloatcreate{1}{1.0}{0}
                             \let\ONE=\pgfmathresult
                             \ifx\result\ONE
                                 \pgfkeysalso{/pgfplots/small value}
                             \else
                                 \pgfkeysalso{/pgfplots/large value}
                             \fi
                         }
                     },
                     check for zero
                 }
             },
             nodes near coords black white=0.24]

                \addplot[matrix plot*, point meta=explicit] file
                 {data/CM_model.csv};
            \end{axis}
        \end{tikzpicture}}
    \quad
    \subfloat[MLP-LDA]{
        \begin{tikzpicture}
            \begin{axis}[
             xlabel={Predictions},
             xticklabels={Winner, Loser, Stabilizer}, xtick={0, 1, 2}, xtick style={draw=none},
             yticklabels={}, ytick style={draw=none},
             width=0.32\textwidth, height=0.32\textwidth,
             enlargelimits=false, axis on top,
             colormap={summap}{color=(white) color=(custom_blue)},
             nodes near coords, nodes near coords align={center},
             nodes near coords black white/.style={
                 small value/.style={text=black},
                 large value/.style={text=white},
                 every node near coord/.style={
                     /pgf/number format/.cd, precision=3, fixed, fixed zerofill, /tikz/.cd,
                     check for zero/.code={
                         \pgfmathfloatifflags{\pgfplotspointmeta}{0}{\pgfkeys{/tikz/coordinate}}{
                             \begingroup
                                 \pgfkeys{/pgf/fpu}
                                 \pgfmathparse{\pgfplotspointmeta<#1}
                                 \global\let\result=\pgfmathresult
                             \endgroup
                             \pgfmathfloatcreate{1}{1.0}{0}
                             \let\ONE=\pgfmathresult
                             \ifx\result\ONE
                                 \pgfkeysalso{/pgfplots/small value}
                             \else
                                 \pgfkeysalso{/pgfplots/large value}
                             \fi
                         }
                     },
                     check for zero
                 }
             },
             nodes near coords black white=0.24]

                \addplot[matrix plot*, point meta=explicit] file
                 {data/CM_meta_model.csv};
            \end{axis}
        \end{tikzpicture}}

    \caption{VA modeling results' normalized confusion matrices for our annotations (main ophthalmologist) as well as our approaches to VA prediction with MLP and MLP-LDA based on our WSL classification scheme, respectively. Note, each confusion matrix shows a single, randomly selected run.}
    \label{figure:VA_modeling_results_CM}
\end{figure}

\begin{table}[tb]
    \newcommand{\hpleq}{\hphantom{<{}}}
    \centering

    \begin{tabular}{|P{5.5cm}|P{2cm}|P{2cm}|P{2cm}|P{2cm}|P{2cm}|}
        \hline
        Condition & Significant? & t & p value & Cohen's d \\
        \hline
        \noalign{\vskip 2pt}

        \hline
        AMD, <1y compared to 1-3y  & 1 & $3.09$ & $\hpleq 0.0020$ & $0.30$ \\
        \hline
        AMD, 1-3y compared to 3-6y & 1 & $4.51$ & $<0.0001$ & $0.35$ \\
        \hline
        AMD, 3-6y compared to >6y  & 0 & $0.59$ & $\hpleq 0.5562$ & $0.05$ \\
        \hline
        AMD, <1y compared to >3y   & 1 & $6.49$ & $<0.0001$ & $\mathbf{0.70}$ \\
        \hline
        \noalign{\vskip 2pt}

        \hline
        RVO, <1y compared to 1-3y  & 0 & $0.57$ & $\hpleq 0.5692$ & $0.07$ \\
        \hline
        RVO, 1-3y compared to 3-6y & 1 & $2.56$ & $\hpleq 0.0110$ & $0.34$ \\
        \hline
        RVO, 3-6y compared to >6y  & 0 & $0.52$ & $\hpleq 0.6048$ & $0.11$ \\
        \hline
        RVO, <1y compared to >3y   & 0 & $1.89$ & $\hpleq 0.0607$ & $0.38$ \\
        \hline
        \noalign{\vskip 2pt}

        \hline
        DME, <1y compared to 1-3y  & 0 & $0.45$ & $\hpleq 0.6500$ & $0.05$ \\
        \hline
        DME, 1-3y compared to 3-6y & 1 & $4.98$ & $<0.0001$ & $0.45$ \\
        \hline
        DME, 3-6y compared to >6y  & 0 & $0.79$ & $\hpleq 0.4277$ & $0.09$ \\
        \hline
        DME, <1y compared to >3y   & 1 & $3.19$ & $\hpleq 0.0016$ & $0.40$ \\
        \hline
        \noalign{\vskip 2pt}

        \hline
        DME compared to DME+ERM, <1y  & 0 & $0.54$ & $\hpleq 0.5860$ & $0.15$ \\
        \hline
        DME compared to DME+ERM, 1-3y & 0 & $0.36$ & $\hpleq 0.7178$ & $0.06$ \\
        \hline
        DME compared to DME+ERM, 3-6y & 0 & $0.12$ & $\hpleq 0.9002$ & $0.02$ \\
        \hline
        DME compared to DME+ERM, >6y  & 0 & $1.16$ & $\hpleq 0.2477$ & $0.32$ \\
        \hline
    \end{tabular}

    \caption{Full statistical test results. See Table~\ref{tab:stat_test_results} for details.}
    \label{tab:stat_test_results_extensive}
\end{table}

\begin{sidewaystable}[tbp]
    \centering

    \begin{tabular}{|M{2cm}|M{2.5cm}|M{2cm}|M{2cm}|M{2cm}|M{2cm}|M{2cm}|M{2cm}|M{2cm}|}
        \multicolumn{9}{c}{\textbf{Ophthalmologist}} \\
        \noalign{\vskip 2pt}

        \cline{3-5}
        \multicolumn{2}{c}{} & \multicolumn{3}{|c|}{Predictions} & \multicolumn{4}{c}{} \\
        \cline{3-9}
        \multicolumn{2}{c|}{} & Winner & Loser & Stabilizer & Total & Ratio [\%] & Precision & Recall \\
        \hline
        \multirow{3}{*}{Ground truth} & Winner & $30$ & $\hphantom{0}2$ & $\hphantom{0\,}110$ & $\hphantom{0\,}142$ & $\hphantom{0}9.5$ & $51.7$ & $21.1$ \\
        \cline{2-9}
        & Loser & $\hphantom{0}3$ & $76$ & $\hphantom{0\,}101$ & $\hphantom{0\,}180$ & $12.1$ & $73.8$ & $42.2$ \\
        \cline{2-9}
        & Stabilizer & $25$ & $25$ & $1122$ & $1\,172$ & $78.5$ & $84.2$ & $95.7$ \\
        \hline
        \noalign{\vskip 2pt}
        \cline{2-9}
        \multicolumn{1}{c|}{} & True positives & \multicolumn{3}{c|}{$1\,228$ (\textbf{$\mathbf{82.2}$~\%})} & $1\,494$ & Mean & $66.3$ & $54.2$ \\
        \cline{2-9}
        \noalign{\vskip 2pt}
        \cline{6-9}
        \multicolumn{5}{c|}{} & \multicolumn{2}{c|}{Macro average F1-score} & \multicolumn{2}{c|}{$57.8$} \\
        \cline{6-9}
        \noalign{\vskip 22pt}

        \multicolumn{9}{c}{\textbf{MLP}} \\
        \noalign{\vskip 2pt}

        \cline{3-5}
        \multicolumn{2}{c}{} & \multicolumn{3}{|c|}{Predictions} & \multicolumn{4}{c}{} \\
        \cline{3-9}
        \multicolumn{2}{c|}{} & Winner & Loser & Stabilizer & Total & Ratio [\%] & Precision & Recall \\
        \hline
        \multirow{3}{*}{Ground truth} & Winner & $\hphantom{0}45$ & $\hphantom{0}10$ & $\hphantom{0}87$ & $\hphantom{0\,}142$ & $\hphantom{0}9.5$ & $22.1$ & $31.7$ \\
        \cline{2-9}
        & Loser & $\hphantom{00}3$ & $\hphantom{0}99$ & $\hphantom{0}78$ & $\hphantom{0\,}180$ & $12.1$ & $28.4$ & $55.0$ \\
        \cline{2-9}
        & Stabilizer & $156$ & $240$ & $776$ & $1\,172$ & $78.5$ & $82.5$ & $66.2$ \\
        \hline
        \noalign{\vskip 2pt}
        \cline{2-9}
        \multicolumn{1}{c|}{} & True positives & \multicolumn{3}{c|}{$920$ ($61.6$~\%)} & $1\,494$ & Mean & $45.8$ & $46.7$ \\
        \cline{2-9}
        \noalign{\vskip 2pt}
        \cline{6-9}
        \multicolumn{5}{c|}{} & \multicolumn{2}{c|}{Macro average F1-score} & \multicolumn{2}{c|}{$45.6$} \\
        \cline{6-9}
        \noalign{\vskip 22pt}

        \multicolumn{9}{c}{\textbf{MLP-LDA}} \\
        \noalign{\vskip 2pt}

        \cline{3-5}
        \multicolumn{2}{c}{} & \multicolumn{3}{|c|}{Predictions} & \multicolumn{4}{c}{} \\
        \cline{3-9}
        \multicolumn{2}{c|}{} & Winner & Loser & Stabilizer & Total & Ratio [\%] & Precision & Recall \\
        \hline
        \multirow{3}{*}{Ground truth} & Winner & $70$ & $\hphantom{0}0$ & $25$ & $\hphantom{0}95$ & $33.3$ & $87.5$ & $73.7$ \\
        \cline{2-9}
        & Loser & $\hphantom{0}0$ & $80$ & $15$ & $\hphantom{0}95$ & $33.3$ & $84.2$ & $84.2$ \\
        \cline{2-9}
        & Stabilizer & $10$ & $15$ & $70$ & $\hphantom{0}95$ & $33.3$ & $63.6$ & $73.7$ \\
        \hline
        \noalign{\vskip 2pt}
        \cline{2-9}
        \multicolumn{1}{c|}{} & True positives & \multicolumn{3}{c|}{$220$ ($77.2$~\%)} & $285$ & Mean & $78.4$ & $77.2$ \\
        \cline{2-9}
        \noalign{\vskip 2pt}
        \cline{6-9}
        \multicolumn{5}{c|}{} & \multicolumn{2}{c|}{Macro average F1-score} & \multicolumn{2}{c|}{$\mathbf{77.5}$} \\
        \cline{6-9}
    \end{tabular}

    \caption{VA modeling results overview of the ophthalmologist, MLP, and MLP-LDA. MLP and MLP-LDA operate with additional data of our completed OCT biomarker documentations. Note, these values show each a single run.}
    \label{table:VA_modeling_results_extensive}
\end{sidewaystable}

\begin{table}[tb]
    \centering

    \subfloat[]{
        \begin{tabular}{|M{2cm}|M{2cm}|M{2cm}|M{2cm}|M{2cm}|}
            \cline{2-4}
            \multicolumn{1}{c|}{} & Precision & Recall & F1-score \\
            \cline{2-4}
            \noalign{\vskip 2pt}

            \hline
            Winner     & $27.3 \pm 17.6$ & $23.7 \pm 16.0$ & $22.3 \pm 14.4$ \\
            \hline
            Loser      & $47.6 \pm 25.7$ & $49.0 \pm 24.9$ & $46.0 \pm 24.6$ \\
            \hline
            Stabilizer & $83.6 \pm \hphantom{0}3.1$ & $80.1 \pm \hphantom{0}8.5$ & $81.2 \pm \hphantom{0}4.2$ \\
            \hline
            \noalign{\vskip 2pt}

            \cline{2-4}
            \multicolumn{1}{c|}{} & \multicolumn{2}{c|}{True positives} & $70.1 \pm \hphantom{0}5.9$ \\
            \cline{2-4}
            \multicolumn{1}{c|}{} & \multicolumn{2}{c|}{Macro average F1-score} & $50.0 \pm 10.7$ \\
            \cline{2-4}
            \noalign{\vskip 2pt}
        \end{tabular}}

    \subfloat[]{
        \begin{tikzpicture}[
         r/.style={draw, rectangle, minimum size=1cm},
         rr/.style={r, align=center, font=\footnotesize}]
            \begin{scope}[shift={(-5, -0.5)}]



                \node[rr, fill=custom_blue!24] at (0,  0) {0.24 $\pm$ \\ 0.16};
                \node[rr, fill=custom_blue!17] at (1,  0) {0.17 $\pm$ \\ 0.19};
                \node[rr, fill=custom_blue!59] at (2,  0) {0.59 $\pm$ \\ 0.23};
                \node[rr, fill=custom_blue!11] at (0, -1) {0.11 $\pm$ \\ 0.16};
                \node[rr, fill=custom_blue!49] at (1, -1) {0.49 $\pm$ \\ 0.25};
                \node[rr, fill=custom_blue!41] at (2, -1) {0.41 $\pm$ \\ 0.24};
                \node[rr, fill=custom_blue!10] at (0, -2) {0.10 $\pm$ \\ 0.06};
                \node[rr, fill=custom_blue!09] at (1, -2) {0.09 $\pm$ \\ 0.08};
                \node[rr, fill=custom_blue!81] at (2, -2) {0.81 $\pm$ \\ 0.09};

                \node at (-1,  0) {W};
                \node at (-1, -1) {L};
                \node at (-1, -2) {S};
                \node at ( 0, -3) {W};
                \node at ( 1, -3) {L};
                \node at ( 2, -3) {S};

                \node[rotate=90] at (-1.5, -1)   {Ground truth};
                \node            at ( 1,   -3.5) {Predictions};
            \end{scope}

            \begin{scope}[shift={(0, 0)}, scale=0.5, transform shape]
                \node[r, fill=custom_blue!03] at (0,  0) {3};
                \node[r, fill=custom_blue!02] at (1,  0) {2};
                \node[r, fill=custom_blue!03] at (2,  0) {3};
                \node[r, fill=custom_blue!00] at (0, -1) {0};
                \node[r, fill=custom_blue!05] at (1, -1) {5};
                \node[r, fill=custom_blue!03] at (2, -1) {3};
                \node[r, fill=custom_blue!03] at (0, -2) {3};
                \node[r, fill=custom_blue!08] at (1, -2) {8};
                \node[r, fill=custom_blue!28] at (2, -2) {28};

                \node at (-1,  0) {W};
                \node at (-1, -1) {L};
                \node at (-1, -2) {S};
                \node at ( 0, -3) {W};
                \node at ( 1, -3) {L};
                \node at ( 2, -3) {S};

                \node[rotate=90] at (-1.5, -1)   {Ground truth};
                \node            at ( 1,   -3.5) {Predictions};
            \end{scope}

            \begin{scope}[shift={(3, 0)}, scale=0.5, transform shape]
                \node[r, fill=custom_blue!01] at (0,  0) {1};
                \node[r, fill=custom_blue!01] at (1,  0) {1};
                \node[r, fill=custom_blue!02] at (2,  0) {2};
                \node[r, fill=custom_blue!02] at (0, -1) {2};
                \node[r, fill=custom_blue!01] at (1, -1) {1};
                \node[r, fill=custom_blue!01] at (2, -1) {1};
                \node[r, fill=custom_blue!08] at (0, -2) {8};
                \node[r, fill=custom_blue!02] at (1, -2) {2};
                \node[r, fill=custom_blue!32] at (2, -2) {32};

                \node at (-1,  0) {W};
                \node at (-1, -1) {L};
                \node at (-1, -2) {S};
                \node at ( 0, -3) {W};
                \node at ( 1, -3) {L};
                \node at ( 2, -3) {S};

                \node[rotate=90] at (-1.5, -1)   {Ground truth};
                \node            at ( 1,   -3.5) {Predictions};
            \end{scope}

            \begin{scope}[shift={(6, 0)}, scale=0.5, transform shape]
                \node[r, fill=custom_blue!03] at (0,  0) {3};
                \node[r, fill=custom_blue!00] at (1,  0) {0};
                \node[r, fill=custom_blue!02] at (2,  0) {2};
                \node[r, fill=custom_blue!01] at (0, -1) {1};
                \node[r, fill=custom_blue!05] at (1, -1) {5};
                \node[r, fill=custom_blue!03] at (2, -1) {3};
                \node[r, fill=custom_blue!03] at (0, -2) {3};
                \node[r, fill=custom_blue!04] at (1, -2) {4};
                \node[r, fill=custom_blue!23] at (2, -2) {23};

                \node at (-1,  0) {W};
                \node at (-1, -1) {L};
                \node at (-1, -2) {S};
                \node at ( 0, -3) {W};
                \node at ( 1, -3) {L};
                \node at ( 2, -3) {S};

                \node[rotate=90] at (-1.5, -1)   {Ground truth};
                \node            at ( 1,   -3.5) {Predictions};
            \end{scope}

            \begin{scope}[shift={(9, 0)}, scale=0.5, transform shape]
                \node[r, fill=custom_blue!01] at (0,  0) {1};
                \node[r, fill=custom_blue!00] at (1,  0) {0};
                \node[r, fill=custom_blue!07] at (2,  0) {7};
                \node[r, fill=custom_blue!00] at (0, -1) {0};
                \node[r, fill=custom_blue!01] at (1, -1) {1};
                \node[r, fill=custom_blue!08] at (2, -1) {8};
                \node[r, fill=custom_blue!08] at (0, -2) {8};
                \node[r, fill=custom_blue!02] at (1, -2) {2};
                \node[r, fill=custom_blue!69] at (2, -2) {69};

                \node at (-1,  0) {W};
                \node at (-1, -1) {L};
                \node at (-1, -2) {S};
                \node at ( 0, -3) {W};
                \node at ( 1, -3) {L};
                \node at ( 2, -3) {S};

                \node[rotate=90] at (-1.5, -1)   {Ground truth};
                \node            at ( 1,   -3.5) {Predictions};
            \end{scope}

            \begin{scope}[shift={(0, -3)}, scale=0.5, transform shape]
                \node[r, fill=custom_blue!01] at (0,  0) {1};
                \node[r, fill=custom_blue!03] at (1,  0) {3};
                \node[r, fill=custom_blue!02] at (2,  0) {2};
                \node[r, fill=custom_blue!01] at (0, -1) {1};
                \node[r, fill=custom_blue!06] at (1, -1) {6};
                \node[r, fill=custom_blue!09] at (2, -1) {9};
                \node[r, fill=custom_blue!16] at (0, -2) {16};
                \node[r, fill=custom_blue!07] at (1, -2) {7};
                \node[r, fill=custom_blue!60] at (2, -2) {60};

                \node at (-1,  0) {W};
                \node at (-1, -1) {L};
                \node at (-1, -2) {S};
                \node at ( 0, -3) {W};
                \node at ( 1, -3) {L};
                \node at ( 2, -3) {S};

                \node[rotate=90] at (-1.5, -1)   {Ground truth};
                \node            at ( 1,   -3.5) {Predictions};
            \end{scope}

            \begin{scope}[shift={(3, -3)}, scale=0.5, transform shape]
                \node[r, fill=custom_blue!01] at (0,  0) {1};
                \node[r, fill=custom_blue!00] at (1,  0) {0};
                \node[r, fill=custom_blue!08] at (2,  0) {8};
                \node[r, fill=custom_blue!00] at (0, -1) {0};
                \node[r, fill=custom_blue!01] at (1, -1) {1};
                \node[r, fill=custom_blue!00] at (2, -1) {0};
                \node[r, fill=custom_blue!01] at (0, -2) {1};
                \node[r, fill=custom_blue!00] at (1, -2) {0};
                \node[r, fill=custom_blue!40] at (2, -2) {40};

                \node at (-1,  0) {W};
                \node at (-1, -1) {L};
                \node at (-1, -2) {S};
                \node at ( 0, -3) {W};
                \node at ( 1, -3) {L};
                \node at ( 2, -3) {S};

                \node[rotate=90] at (-1.5, -1)   {Ground truth};
                \node            at ( 1,   -3.5) {Predictions};
            \end{scope}

            \begin{scope}[shift={(6, -3)}, scale=0.5, transform shape]
                \node[r, fill=custom_blue!01] at (0,  0) {1};
                \node[r, fill=custom_blue!03] at (1,  0) {3};
                \node[r, fill=custom_blue!04] at (2,  0) {4};
                \node[r, fill=custom_blue!00] at (0, -1) {0};
                \node[r, fill=custom_blue!03] at (1, -1) {3};
                \node[r, fill=custom_blue!03] at (2, -1) {3};
                \node[r, fill=custom_blue!02] at (0, -2) {2};
                \node[r, fill=custom_blue!08] at (1, -2) {8};
                \node[r, fill=custom_blue!31] at (2, -2) {31};

                \node at (-1,  0) {W};
                \node at (-1, -1) {L};
                \node at (-1, -2) {S};
                \node at ( 0, -3) {W};
                \node at ( 1, -3) {L};
                \node at ( 2, -3) {S};

                \node[rotate=90] at (-1.5, -1)   {Ground truth};
                \node            at ( 1,   -3.5) {Predictions};
            \end{scope}

            \begin{scope}[shift={(9, -3)}, scale=0.5, transform shape]
                \node[r, fill=custom_blue!01] at (0,  0) {1};
                \node[r, fill=custom_blue!00] at (1,  0) {0};
                \node[r, fill=custom_blue!06] at (2,  0) {6};
                \node[r, fill=custom_blue!01] at (0, -1) {1};
                \node[r, fill=custom_blue!03] at (1, -1) {3};
                \node[r, fill=custom_blue!02] at (2, -1) {2};
                \node[r, fill=custom_blue!04] at (0, -2) {4};
                \node[r, fill=custom_blue!01] at (1, -2) {1};
                \node[r, fill=custom_blue!34] at (2, -2) {34};

                \node at (-1,  0) {W};
                \node at (-1, -1) {L};
                \node at (-1, -2) {S};
                \node at ( 0, -3) {W};
                \node at ( 1, -3) {L};
                \node at ( 2, -3) {S};

                \node[rotate=90] at (-1.5, -1)   {Ground truth};
                \node            at ( 1,   -3.5) {Predictions};
            \end{scope}
        \end{tikzpicture}}

    \caption{VA modeling results overview of our control experiment with eight different additional ophthalmic doctors given randomized subsets of our test set. \textbf{(a)} Precision, recall, and F1-score comparison for our WSL-based classes. \textbf{(b)} Confusion matrices, normalized via mean and standard deviations over all doctors (left), and single confusion matrices for each doctor with absolute results (right).}
    \label{table:VA_modeling_results_extensive_8_doctors}
\end{table}

\subsection{Discussion}
\label{section:discussion}

\textit{Better performance due to the completion of OCT biomarkers.} In our work, we developed a multistage system that completes previously incomplete OCT biomarker documentations by means of learning-based approaches, which are then utilized for the following visual acuity prediction. This approach enabled us to provide additional data that were previously not available to be included into the AI modeling process, therefore achieving an improved model performance. The OCT biomarker classification recognizes OCT biomarkers based on the OCT B-scan images for patients where OCT biomarkers were previously not available within the electronic health records, resulting in completed OCT biomarkers (section~2.5.3). Subsequently, these completed OCT biomarkers documentations are exploited in our MLP-LDA system, together with the already existing OCT biomarker information, the visual acuity, and other medical data, to predict the visual acuity (section~2.5.4). The ophthalmologists had therefore only access to the already existing OCT biomarker documentations. They could only utilize these data. Moreover, in larger hospitals, different ophthalmologists are included in the diagnosis. They often have to rely on previous documentations, including previous OCT biomarker diagnoses. In addition, fellow-level ophthalmologists are often included in the diagnosis to create OCT biomarker documentations. Thus, even the existing data might not have the best quality. Some specialists work around these weaknesses by directly analyzing the OCT B-scan images, too. We have performed control experiments to conduct the magnitude of the influence effect from our novel, completed OCT biomarker documentations. The full system achieves a performance of $69.0$~\% F1-score (Table~6). When excluding the completed OCT biomarker documentations, we obtain an accuracy of $62.8$~\%. Hence, the OCT biomarker completion is valid for a noticeable improvement of ca. $+6$~\% F1-score. Within our (semi-)automated context, we therefore see the ability to complete OCT biomarker documentations as a major benefit of our system. Finally, out of over 49\,000 patients with overall more than 130\,000 examinations, we identified approximately 1\,500 AMD/DME/RVO patients with about 15\,000 relevant examinations. It can be assumed that a learning-based approach such as ours leverages the latent knowledge within our data corpus.

\textit{Differences in therapy winner/stabilizer/loser classification.} In Table~8, it is shown that our test set consists of mostly patients with progressions of the category therapy stabilizers given the visual acuity values of adjacent time steps (ca. 80~\%). For this reason, the MLP model has learned that at average a stabilizer is to be expected. Therefore, most classifications for therapy winners and losers are incorrectly assigned to the class of therapy stabilizers. Analogous, our trained ophthalmologist knows of this general distribution within our classification scheme. The MLP-LDA approach has learned a more fine-grained classification than MLP due to the subsequent correction/control stage, for which we observe a shift from mostly classified therapy stabilizers towards therapy winners and losers (Fig.~5 and Table~6). This allows us to achieve a generally improved classification performance in macro average F1-score, which explains the mechanism and possible benefits of our correction stage using MLP-LDA. Subsequently, all ophthalmologists show a sizeable advantage in classifying stabilizers with improved precision, recall, and F1-scores (Tables~8 and 9). In comparison to MLP, it is evident that MLP-LDA excels at differentiating between therapy stabilizers as well as winners and losers, demonstrated by an improvement in macro average F1-score and class-wise precision/recall. Future work could thus focus on therapy loser modeling as therapy losers are the relevant subgroup where a better treatment handling would be important. However, this requires that further therapy options can be reliably modeled for this group to recommend a potentially better therapy option. As mostly therapy stabilizers are present, the ophthalmologist obtains the best accuracy in true positives with $82.2$~\% in comparison to MLP-LDA with $77.2$~\% (see also Table~8). The other eight ophthalmologists score $70.1 \pm 5.9$~\% in true positives, which signals results within the same range. We conclude that, depending on the evaluated progression group as well as the related evaluation metrics, different advantages and disadvantages in prediction accuracies can be observed.

\textit{The influence of medical features on the visual acuity prediction accuracy.} To understand the influence of single components, we have performed experiments to include or to omit particular medical features. In Table~6, we have chosen a 3-step approach. Firstly, we started with an input of visual acuity values only to our system, where a performance of $40$--$45$~\% F1-score shows that time and visual acuity history already have a certain prediction potential. To give an idea of the scale, the annotating ophthalmologists achieve $57.8$~\% and $50 \pm 10.7$~\% F1-score given a constant IVOM therapy scheme. Secondly, we added all medical features (cf. Table~3c) to our models, whereby two of our models, MLP and MLP-LDA, benefited from it distinctly. The performances led to a score of about $45$~\% and $69$~\% F1-score, respectively. In the last step, we omitted the OCT biomarker documentations, which resulted in a decrease in prediction accuracy of a few percentage points, thus highlighting the influences of OCT biomarkers. The analyses were conducted separately for the existing OCT biomarkers and for the completed OCT biomarkers (Table~6). The omission/inclusion of OCT biomarkers in combination with the tracking of the evaluation metrics is aiding in quantifying their predictive potentials. However, we did not yet explore the full potential of biomarkers such as DRIL and hyperreflective materials, which are relevant for DME. This will be a part of our future work. The omission/inclusion of particular medical features is similarly relevant for understanding the cues of medical features, and by means of AI technology, deepen the understanding of the field of ophthalmology.

\textit{Differences between the AI and ophthalmologists.} Although we have tried to create the same starting conditions for the comparison of ophthalmologists and AI, this is not always achievable. In the following, we highlight some of the occurring difficulties and differences. The proposed learning-based system completes OCT biomarkers during its evaluation process, whereas the ophthalmologist does not have access to them. For a fairer comparison, we have to compare the performance of MLP-LDA without the completed OCT biomarkers ($62.8$~\% macro average F1-score) with the ophthalmologists ($58$ and $50 \pm 10.7$~\% F1-score). This is possible since we have also benchmarked MLP-LDA without the completed OCT biomarkers (see also Table~6). Furthermore, the ophthalmologists were not trained on the same number of data vectors as our AI system beforehand, which was trained on about 1\,200 patients with at least five visual acuity examinations (80~\% train set out of 1\,496 total patients). A daily clinical routine might provide some of the highly-experienced ophthalmologists with such numbers of patients, while freshly-educated ophthalmologists might not have seen such a number of patients yet. Our ophthalmologists have different experience and education levels. On the other hand, the ophthalmologists have a medical school or university study over several years, which the AI system, logically, does not have. Patient data in the dashboard view were not provided to them for training but instead only for their visual acuity prediction. As the training levels of the ophthalmologists are heterogeneous, their varying results were also expected to vary in addition to the already expected variety when working with human subjects, accordingly. AI models and ophthalmologists both had to deal with data incompleteness. Humans may experience high performance variability even on the same task in behavioral experiments depending on factors such as their form on the day, attentional state in their mind, motivational level, and mental workload. Finally, the ophthalmologists do have background knowledge on how to interpret time series data. For example, they could calculate or estimate the duration of diabetes from DME diagnoses as a risk factor, which was not directly provided in our case as we provided some information but no duration of diabetes. For longer time series, this knowledge could prove to be an advantage for ophthalmologists. However, background knowledge might also bias ophthalmologists to predict more stabilizers, assuming the distribution within our test set, whereby stabilizers are the largest progression group for our experimental setting (comparing $t$ and $t - 1$ to determine the resulting WSL classification). However, no ophthalmologist was informed about this distribution within our test set.

\textit{Explainable AI and training of human raters.} To quote from ``Explainable Artificial Intelligence (XAI): Concepts, taxonomies, opportunities and challenges toward responsible AI'' by \textit{Arrieta et al.}: ``... in Machine Learning, the entire community stands in front of the barrier of explainability, an inherent problem of the latest techniques brought by sub-symbolism (e.g. ensembles or Deep Neural Networks) that were not present in the last hype of AI (namely, expert systems and rule based models).'' \cite{arrieta2020explainable}. We understand that comprehensibility could foster theoretical improvements in the field, which requires to represent our models' learned knowledge in a human-understandable fashion. However, in the context of MLP and LDA as well as high-dimensional data vectors, we believe this to be a distinct effort, which is not trivial in nature, hence beyond the scope of this work. A training of physicians by using the cues that the models use is therefore not possible at the moment. Even decisions trees who give comprehensible rules can be too complex to derive straightforward rules given high-dimensional data such as ours. An approach to determine the possible influence of single components, i.e., medical features, via their omission/inclusion has been performed in our experiments (section~3.3), which gives some comprehensibility on levels. For the image classification of OCT biomarkers using deep neural networks on the other hand, a gradient-weighted class activation mapping (Grad-CAM) could be applied \cite{selvaraju2016grad, selvaraju2017grad} to reason the regions of interest within the OCT images where neurons were especially active. This will be a part of our future studies. However, the aggregated visualization within a dashboard such as ours could be already utilized to train physician (i) with more patient progressions, (ii) more specific information such as the present type of a disease or treatment, or (iii) lengthy progressions that would be rarely seen otherwise.

\section{Conclusion and outlook}
\label{section:conclusion_and_outlook}

In this contribution, we developed an IT system architecture that aggregates patient-wise information for more than 49\,000 patients from different categories of various multimedia data in the form of text and images within multiple heterogeneous ophthalmic data resources originating from a German hospital of maximum care. As the prediction of a patient's progression is challenging within this real-world setting, our realized workflow allows a first processing of medical patient data to enable an OCT biomarker classification, a visual acuity prediction, as well as a general statistical evaluation and visualization. For this purpose, our developed patient progression visualization and modeling dashboard enables the visualization, annotation, and assessment of patient progressions with a focus on their visual acuity.

The resulting data corpus allows predictive statements of the expected progression of a patient and his or her visual acuity in each of the three diseases AMD, DME, and RVO. Our data reveal that especially exudative AMD results in a notable high amount of therapy ``losers'' ($60$~\% regarding a time span of 3 to 6 years). The result for AMD is significant. Furthermore, we found a weakly significant deterioration of visual acuity for DME, while we found no significant deterioration for RVO. A more fine-grained analysis is able to reveal the influence of medical co-existing factors such as other diseases. As a proof of concept, we exemplary show DME with an epiretinal membrane. Yet, the data situation is still too weak to derive reliable correlations for statistical surveys of comorbidities in combination with different observation time windows.

For the following visual acuity based treatment progression modeling, incomplete OCT documentations are completed by classifying the OCT scans' slices (OCT B-scans), which in turn allows the classification of OCT scans when only single OCT slices are provided. Based on the obtained OCT slice classifications, a scan-wise OCT classification of the OCT biomarkers ELM, ellipsoid zone, foveal depression, RPE, scars, and subretinal fibrosis resulted in an overall classification accuracy of over $98$~\% in F1-score. Finally, the completed OCT documentations are combined with additional medical data, defining our ophthalmic feature vectors for visual acuity prediction. In comparison to different approaches from machine learning and deep learning, we achieve a final prediction accuracy of $69$~\% in macro average F1-score with $77.2$~\% true positives, while our main ophthalmologist shows a macro average F1-score of $57.8$~\% with $82.2$~\% true positives. In order to further validate these results, we evaluated the annotations of eight different additional ophthalmic doctors given randomized subsets of our test set, resulting in an overall macro average F1-score of $50.0 \pm 10.7$~\% and with $70.1 \pm 5.9$~\% true positives.

However, as the influence of the OCT biomarkers is not yet fully understood, further investigations have to be conducted, for which additional OCT biomarkers as well as their influence for the visual acuity modeling process have to be evaluated. Future contributions can build on these initial results in order to determine an optimal time for a change in medication or therapy. This also encompasses treatment options such as laser coagulation, pars plana vitrectomy, or phacoemulsification with posterior chamber lens implantation. We furthermore aim at extending our approach to include a larger data corpus through distributed analysis across multiple ophthalmic sites. Thus, data quality needs to be ensured via comprehensive evaluations of our medical texts structured by rule- and learning-based NLP methods, which requires further harmonization of the underlying medical terminology. Patient-related data of the different categories available and their relevance for the modeling process have to be further investigated in order to increase the evidence of AI-based modeling approaches to enable future realizations of a (semi-)automated recommender system.

\section*{Acknowledgment}

This research was partially funded by the European Social Fund for Germany, the Novartis Pharma GmbH, as well as the German Federal Ministry of Education and Research, namely the \textit{Medical Informatics Hub in Saxony (MiHUBx)} with the grant numbers 01ZZ2101E (Klinikum Chemnitz) and 01ZZ2101C (Chemnitz University of Technology).

\section*{Author contributions}

Tobias Schlosser, Frederik Beuth, Trixy Meyer, and Danny Kowerko conducted this contribution's writing process and the related research project's implementation and evaluation with the help of Arunodhayan Sampath Kumar, Gabriel Stolze, Olga Furashova, and Katrin Engelmann in revising this manuscript, whereas Olga Furashova, Katrin Engelmann, and Danny Kowerko designed and supervised the related research project.

\section*{Competing interests}

The authors declare that they have no conflict of interest.

\section*{Data availability}

The data that support the findings of this study are available from the Department of Ophthalmology at the Klinikum Chemnitz gGmbH in Chemnitz, Germany but restrictions apply to the availability of these data, which were used under license for the current study, and so are not publicly available. Data are however available from the authors upon reasonable request and with permission of the Data Integration Center of the Klinikum Chemnitz gGmbH. Inquiries on this matter should be addressed to Danny Kowerko (danny.kowerko@cs.tu-chemnitz.de).

\bibliography{library}

\end{document}